\documentclass[conference]{IEEEtran}

\ifCLASSINFOpdf
\else
\fi
\usepackage{algorithm}
\usepackage{algpseudocode}
\usepackage{hyperref}
\usepackage{caption}
\usepackage{subcaption}
\usepackage{amsmath}
\usepackage{amssymb}
\usepackage{graphicx}
\usepackage{soul}

\usepackage{mdframed}
\usepackage{enumitem}
\usepackage{multirow}
\usepackage{xcolor}
\usepackage[normalem]{ulem}
\usepackage{balance}
\newcommand{\etal}{\textit{et al.}}
\definecolor{customgreen}{HTML}{C5DCA0}
\definecolor{customblue}{HTML}{90e0ef}
\definecolor{transitionred}{HTML}{ff0000}
\definecolor{transitionorange}{HTML}{f8961e}
\definecolor{transitionpink}{HTML}{ff70a6}
\begin{document}
%
\title{\textit{Bit of a Close Talker}: A Practical Guide to Serverless Cloud Co-Location Attacks}

\author{
\IEEEauthorblockN{
    Wei Shao\IEEEauthorrefmark{1},
    Najmeh Nazari\IEEEauthorrefmark{1},
    Behnam Omidi\IEEEauthorrefmark{2},
    Setareh Rafatirad\IEEEauthorrefmark{1},\\
    Khaled N. Khasawneh\IEEEauthorrefmark{2},
    Houman Homayoun\IEEEauthorrefmark{1}, and
    Chongzhou Fang\IEEEauthorrefmark{3}
}
\IEEEauthorblockA{\IEEEauthorrefmark{1}University of California, Davis, USA\\
Email: \{wayshao, nnazari, srafatirad, hhomayoun\}@ucdavis.edu}
\IEEEauthorblockA{\IEEEauthorrefmark{2}George Mason University, USA\\
Email: \{bomidi,kkhasawn\}@gmu.edu}
\IEEEauthorblockA{\IEEEauthorrefmark{3}Rochester Institute of Technology, USA\\
Email: cxfeec@rit.edu}
}
	

%


\IEEEoverridecommandlockouts
\makeatletter\def\@IEEEpubidpullup{6.5\baselineskip}\makeatother
\IEEEpubid{\parbox{\columnwidth}{
		Network and Distributed System Security (NDSS) Symposium 2026\\
		23 - 27 February 2026 , San Diego, CA, USA\\
		ISBN 979-8-9919276-8-0\\  
		https://dx.doi.org/10.14722/ndss.2026.241376\\
		www.ndss-symposium.org
}
\hspace{\columnsep}\makebox[\columnwidth]{}}

\maketitle

\begin{abstract}
Serverless computing has revolutionized cloud computing by offering users an efficient, cost-effective way to develop and deploy applications without managing infrastructure details. However, serverless cloud users remain vulnerable to various types of attacks, including micro-architectural side-channel attacks. These attacks typically rely on the physical co-location of victim and attacker instances, and attackers need to exploit cloud schedulers to achieve co-location with victims. Therefore, it is crucial to study vulnerabilities in serverless cloud schedulers and assess the security of different serverless scheduling algorithms. This study addresses the gap in understanding and constructing co-location attacks in serverless clouds. We present a comprehensive methodology to uncover exploitable features in serverless scheduling algorithms and to devise strategies for constructing co-location attacks via normal user interfaces. In our experiments, we successfully reveal exploitable vulnerabilities and achieve instance co-location on prevalent open-source infrastructures and Microsoft Azure Functions. We also present a mitigation strategy, the
\textit{Double-Dip} scheduler, to defend against co-location attacks in serverless clouds. Our work highlights critical areas for security enhancements in current cloud schedulers, offering insights to fortify serverless computing environments against potential co-location attacks.
\end{abstract}


%
\IEEEpeerreviewmaketitle

\section{Introduction}
In recent years, serverless computing, a relatively new form of cloud computing, has become prevalent and attracts a great number of users. It fully offloads the task of infrastructure management to cloud providers, enabling developers to focus more on the functionality of their applications and to host services more cost-effectively~\cite {shafiei2022serverless,li2022serverless}. Today, we can still see an increasing number of commercial services pivoting to this cloud computing model. This computing paradigm has been supported by mainstream cloud providers, e.g., Amazon AWS Lambda~\cite{awslambda}, Google Cloud Run~\cite{googlecloudrun}, and Microsoft Azure Functions~\cite{microsoftazurefunctions}. Compared to traditional cloud computing (often referred to as ``serverful computing''), serverless computing has the following benefits: (1) Flexibility, as the cloud provider-managed light-weight serverless instances can flexibly scale depending on the load of clients' services and cloud providers can achieve higher resource efficiency; (2) Convenience, since client developers now only need to develop functional code and directly submit code or container images to cloud providers, with all resource management details being 
hidden from them; (3) Customer cost efficiency, as customers are only charged for the actual resources they utilize (often measured by the number of function invocations), unlike in serverful cloud computing, where customers are charged by instance time and a lot of costs are wasted on instance idling.

Resource provisioning, i.e., scheduling, is an important aspect of serverless cloud operation. It impacts the performance of customer applications and can harm the profits of cloud providers if serverless instances are not properly placed and provisioned. Apart from resource scheduling challenges that are similar to serverful clouds, one unique challenge in serverless scheduling is the existence of cold-start~\cite{coldstart}, i.e., the time-consuming start-up process of serverless instances. Another unique challenge arises from the flexible nature of resource management of serverless systems, and serverless cloud providers need to adaptively perform auto-scaling to accommodate the dynamic service workloads. There are lots of research works dedicated to improving the performance of serverless systems from scheduling policies~\cite{fuerst2022locality,li2022help,agarwal2024deep}.

Despite being convenient and cost-effective for cloud users, serverless clouds still face security challenges posed by various types of micro-architectural side-channel attacks. For example, remote micro-architectural attacks like Prime+Probe~\cite{disselkoen2017prime+} remain feasible in serverless clouds with certain optimizations~\cite{zhao2024last}, threatening the security of serverless users. These attacks depend on the co-location of attackers' and victims' serverless instances on the same physical server within the cloud cluster. Although perfect isolation can be obtained by switching to dedicated hosts, this comes at a prohibitive cost: the cheapest dedicated servers from AWS and Azure start at \$323 and \$444 per month, respectively, which is over 10 times higher than running the same workload on serverless platforms, as discussed in Section~\ref{Sec:OtherDefense}. Therefore, beyond understanding the mechanics of the actual attack, it is crucial to study how this critical prerequisite of co-location can be achieved and to analyze potential attacker strategies. This analysis will inspire the design of more secure serverless schedulers. Co-location attack methods have been investigated in traditional clouds~\cite{fang2022repttack,makrani2021cloak,ristenpart2009hey,varadarajan2015placement} and Google Cloud Run~\cite{zhao2024everywhere}. However, there is still a lack of studies on a universal methodology to exploit serverless cloud schedulers.

In this work, we aim to bridge this gap in the literature and provide universal guidance on the construction of co-location attacks targeting different serverless clouds. We first provide a method to reveal exploitable features of serverless cloud scheduling algorithms, such as package locality optimizations~\cite{aumala2019beyond}. Then, based on the found features, we offer a strategy to construct attacks accordingly. All these steps are done through a normal user interface, i.e., besides deploying serverless applications and invoking corresponding services, attackers do not have other access permissions to the cloud infrastructure. The main design effort is hence dedicated to constructing a set of attacks against serverless services from the user side.

The main contributions of this paper are summarized as follows:
\begin{itemize}
    \item We propose a method to analyze and reveal serverless scheduler features;
    \item We develop a strategy for constructing co-location attacks based on the identified scheduler features;
    \item We conduct evaluations on different types of schedulers, uncovering key insights and offering guidance for improving scheduler design;
    \item As a case study, we successfully attack Azure Functions and manage to achieve co-location of serverless instances from different accounts. 
    \item Additionally, we propose a scheduling algorithm called \textit{Double-Dip} that aims to provide defense by offering ``soft'' user isolation.
\end{itemize}

The remainder of this paper is organized as follows. In Section~\ref{SecBackGround}, we provide an introduction to background knowledge like serverless scheduling algorithms. We introduce our scheduler fingerprinting strategy in Section~\ref{SecScheFinp} and evaluate the effectiveness of our method in Section~\ref{SecFinpEval}. Section~\ref{SecAttack} proposes the strategies to construct attacks, and the evaluation results are presented in Section~\ref{SecAttackEval}. The case study on Microsoft Azure Functions is provided in Section~\ref{SecCase}. Our proposed \textit{Double-Dip} scheduler as a mitigation is discussed and evaluated in Section~\ref{SecMit}. Additional discussions are included in Section~\ref{SecDisc}. Finally, a review of the literature and our conclusion are presented in Section~\ref{SecRelatedWork} and Section~\ref{SecConclusion}, respectively.

\section{Background}\label{SecBackGround}
\subsection{Serverless Schedulers}
Serverless cloud offloads all infrastructure-related tasks to cloud providers, making cloud providers fully responsible for managing serverless function instances and properly performing operations like host launching, scaling out, etc. It is hence important to develop scheduling policies that determine the optimal distribution of resources.

Serverless scheduling is different from traditional scheduling in serverful clouds. In serverless clouds, user code is executed in encapsulated environments that pack required dependencies, e.g., containers or vendor-specific instances~\cite{agache2020firecracker}. Serverless schedulers are responsible for deciding the placements of these instances. Like in traditional cloud scheduling and cluster scheduling, serverless schedulers need to determine how resources are assigned to user instances in the system based on resource availability. Additionally, serverless schedulers need to make decisions on how to dispatch serverless function invocations to instances. A major thread of research is dedicated to optimizing start time and avoiding unnecessary cold-start latency by optimizing scheduling policies~\cite{agarwal2021reinforcement,suresh2020ensure,sethi2023lcs}. Besides, as serverless cloud providers are also responsible for dynamically scaling function instances~\cite{azurehostingoptions}, auto-scaling and load-balancing are also important aspects of serverless schedulers. There are research works that consider different kinds of locality~\cite{abdi2023palette,aumala2019beyond} to optimize for load balancing performance. Later in this paper, we will show that from an attacker's view, these are features that can be exploited to increase the co-location attack success rate. 

\subsection{Cloud Co-Location Attacks}
To launch malicious attacks like micro-architectural side-channel attacks~\cite{yarom2014flush+,gruss2016flush+,kocher2019spectre,lipp2018meltdown}, an important prerequisite that attackers must achieve is the co-location of victim and attack instances. Without obtaining co-location, attackers cannot gain access to shared resources like cache and will be unable to launch subsequent attacks. Cloud co-location attacks~\cite{makrani2021cloak, fang2022repttack, fang2023heteroscore, zhao2024everywhere} aim to exploit scheduler features to manipulate instance placement results in cloud systems and force attacker and victim instances to be placed together on the same physical machine. 

\begin{figure}[ht!]
    \centering
    \includegraphics[width=.9\linewidth]{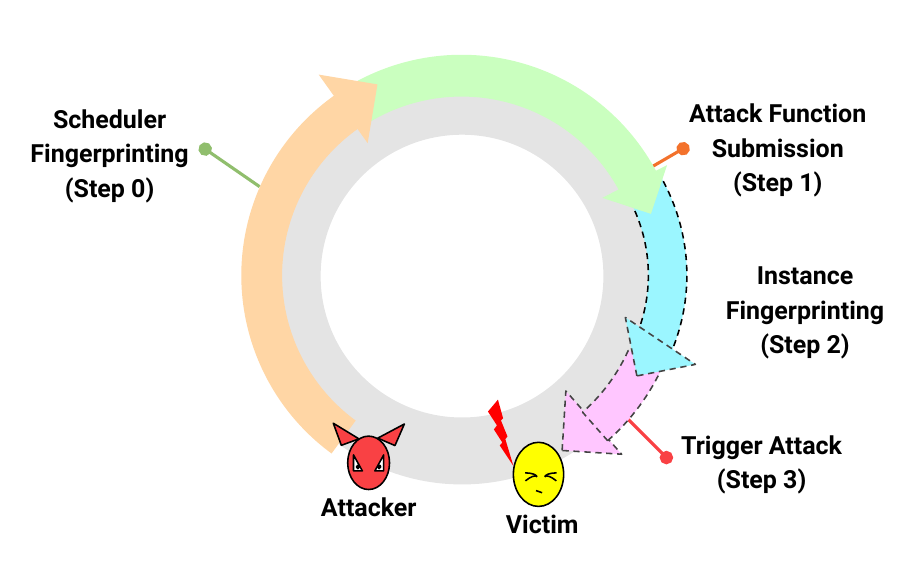}
    \caption{Diagram of the overall attack flow.}
    \label{FigServerlessAttack}
\end{figure}

In both serverful and serverless clouds, a side-channel attacker needs to go through the following steps to eventually complete an attack, as shown in Fig.~\ref{FigServerlessAttack}.

\begin{enumerate}[label=Step \arabic*., start=0, left=0cm]
    \item Use trial submissions to collect information about the scheduler of the target cloud, and optimize submission strategies accordingly;
    \item Invoke attack functions following a predefined policy;
    \item Fingerprint co-residing applications and verify if any attacker instance is co-located with a victim;
    \item Launch attack and steal secret information from the victim.
\end{enumerate}

Step 3 is the focus of most side-channel attack works, while Steps 0 - 2 are usually neglected by researchers. However, achieving co-location is a non-trivial process, especially as the size of datacenters increases over time. Prior works in recent years~\cite{makrani2021cloak,fang2022repttack,zhao2024everywhere} indicate that special strategies should be constructed according to scheduler features to tamper with targeted cloud systems.

\subsection{Definitions}
In the remainder of this paper, user-deployed serverless services are denoted as `functions', while corresponding cloud instances are referred to as `function hosts'. The requests to execute a function in the cloud, regardless of trigger types, are called function invocations. 

\subsection{Threat Model}
In this paper, we only study Step 0 (Section~\ref{SecScheFinp}) - Step 1 (Section~\ref{SecAttack}) of the co-location attack. We assume non-privileged attackers, i.e., attackers can only launch attacks by invoking functions and collecting the returned execution results. The goal of attackers is to place \textbf{at least} one of their attack function hosts together with a victim's function host on the \textbf{same} physical machine to enable subsequent attack efforts. In addition, we assume that the attacker does not have any knowledge about the targeted serverless scheduler. Scheduling strategies need to be reverse-engineered (using our method in Section~\ref{SecScheFinp}), based on which the attacker constructs an attack strategy accordingly to target a specific scheduler. However, they can have certain knowledge about the targeted application, e.g., dependencies, which is realistic in the real world. Examples include Airbnb's disclosure of their usage of open-source projects~\cite{airbnb} and Netflix's publication of their internal tool as an open-source project~\cite{netflix}.

\section{Scheduler Fingerprinting}\label{SecScheFinp}
We first propose a scheduler fingerprinting method to reveal certain important features in the scheduling algorithm designs. This will be the foundation of subsequent attack construction.
\subsection{Algorithm Description}
The goal of scheduler fingerprinting is to identify exploitable features of the scheduling algorithm of the target serverless cloud. The obtained information can be utilized to construct attacks targeting the cloud scheduler. By carefully crafting the function invocation sequence, an attacker will be able to obtain information about the scheduling algorithms based on the collected placement sequences.

In this part, we assume a reliable server identification method is provided, i.e., when invoking functions, the attacker will be able to differentiate servers where the function host is placed. This can be based on methods such as time and frequency probing~\cite{zhao2024everywhere} or network-based probing~\cite{ristenpart2009hey}. With this server fingerprinting method, an attacker can continuously `discover' servers in the cloud system and provide unique identifiers to each discovered physical server. Following the same function invocation sequence, the collected trace of such identifiers will be different for different scheduling algorithms, hence they can be used to identify scheduling policies utilized by the cloud provider. 

We denote the aforementioned server identification method as a function $\mathcal{F}$. Every time a serverless function $\phi$ is executed in the cluster, a hypothetical fingerprint of the corresponding physical server will be reported [denoted as $\mathcal{F}(\phi)$]. We maintain a mapping record ($\mathcal{R}$) between physical server fingerprints and physical server IDs. When a fingerprint that does not map to any server is returned, a new server in the cluster is discovered. A new ID will be assigned to the server, and a new entry in $\mathcal{R}$ will be established to record the mapping between the server fingerprint and the assigned server ID. The trace of server IDs associated with returned server fingerprints after each function invocation is recorded and will later be used for analysis. 
As an example, consider a function that is invoked $6$ times. If the first $3$ executions are handled by one server in the cluster, and the last $3$ executions are on another server, the resulting sequence will be $(1, 1, 1, 2, 2, 2)$.

\begin{figure}[ht!]
    \centering
    \includegraphics[width=\linewidth]{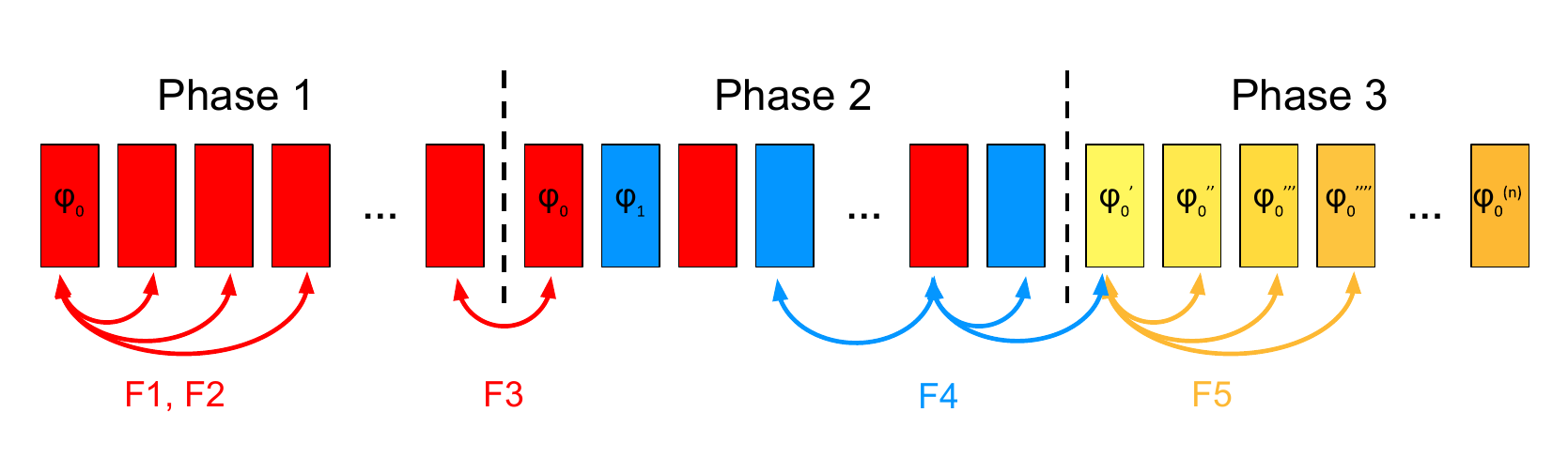}
    \caption{The scheduler fingerprinting process.}
    \label{FigFinp}
\end{figure}

Shown in Fig.~\ref{FigFinp}, the submission of function invocation consists of $3$ phases (a detailed pseudocode description can be found in Appendix~\ref{AppAlg}):

\textbf{Phase 1.} In this step, the attacker constructs a serverless function ($\phi_0$) and continuously triggers it for a specific length of time (e.g., 10 minutes). Then, the attacker will stop all activities and wait for a sufficiently long amount of time (e.g., 1 hour) so that all function hosts are cleared/terminated by the cloud system.

\textbf{Phase 2.} In this step, the attacker constructs a second serverless function $\phi_1$, which is a copy of the previous serverless function $\phi_0$. Then, the attacker will invoke the two functions alternately for a specific length of time (e.g., 10 minutes), and then wait for all function hosts to be terminated.

\textbf{Phase 3.} In this step, the attacker considers possible locality optimization that the cloud provider may use and tests these options. Possible target locality optimization techniques include data locality~\cite{abdi2023palette}, package locality~\cite{aumala2019beyond}, etc. The attacker will construct a list of serverless functions $\{\phi_0', \phi_0'', ... \phi_0^{(n)}\}$ that is functionally identical with $\phi_0$ but vary in locality-related attributes. For example, when targeting package locality, an attacker will construct variants of $\phi_0$ that request different packages to be installed.

\textbf{Analysis.} Features of the scheduling algorithm can be revealed by analyzing and comparing different parts of the collected trace. In Phase 1, the attacker invokes the same function with a relatively high frequency for a sufficiently long pre-defined length of time. The placement sequence in the collected trace can reveal the following features ($\mathrm{F}_i$) of the scheduler:
\begin{enumerate}[label=$\mathrm{F}_{\arabic*}$.]
    \item \textit{Invocation Locality}: If these functions are consistently executed on the same machine over a continuous period, it indicates that the scheduler considers reusing already placed function hosts to avoid cold-start delays.
    \item \textit{Auto-Scaling}: If $\mathrm{F}_1$ is observed and new servers are discovered continuously, it indicates that there are auto-scaling mechanisms embedded in the scheduling algorithm.
\end{enumerate}

In Phase 2, after all function hosts are terminated, the attacker re-invokes the same function in Phase 1, together with another function with the same functionality but disguised as a different function. This enables the attacker to compare placement results between (1) invocations to the same function that are separated in time, and (2) invocations to different functions. The attacker will be able to answer the following questions:
\begin{enumerate}[label=$\mathrm{F}_{\arabic*}$., start=3]
    \item \textit{Cold-Start Locations}: By comparing the placement results of the first function in Phase 1 and Phase 2, the attacker will be able to obtain information regarding the scheduler's behavior when rescheduling a function instance that has been terminated before. The attacker will be able to answer the following question: for the same function, is the cold-start location always the same? 
    \item \textit{Account Locality}: By comparing placement results of different functions, the attacker will be able to infer if user locality is considered by the scheduler and answer this question: do the schedulers tend to place function hosts from the same user on the same set of physical servers? This can potentially infect the construction of an attack strategy. This also needs to be cross-checked with placement results from Phase 3.
\end{enumerate}

The goal of Phase 3 is to explore whether there are other locality optimizations that rely on specific configurations of serverless functions (e.g., package required) and are employed by the scheduler. By comparing the placement results of different functions, the attacker will be able to identify:
\begin{enumerate}[label=$\mathrm{F}_{\arabic*}$., start=5]
    \item \textit{Configuration-Based Locality}: By invoking functions with different configuration parameters (packages, required data, etc.), the attacker will be able to identify certain configuration-based locality optimizations for further exploitation. For example, the scheduler considers package locality and places function hosts with similar package specifications on the same machine.
\end{enumerate}

\section{Fingerprinting Evaluation}\label{SecFinpEval}
\subsection{Simulator Details}
\begin{figure}[ht!]
    \centering
    \includegraphics[width=.8\linewidth]{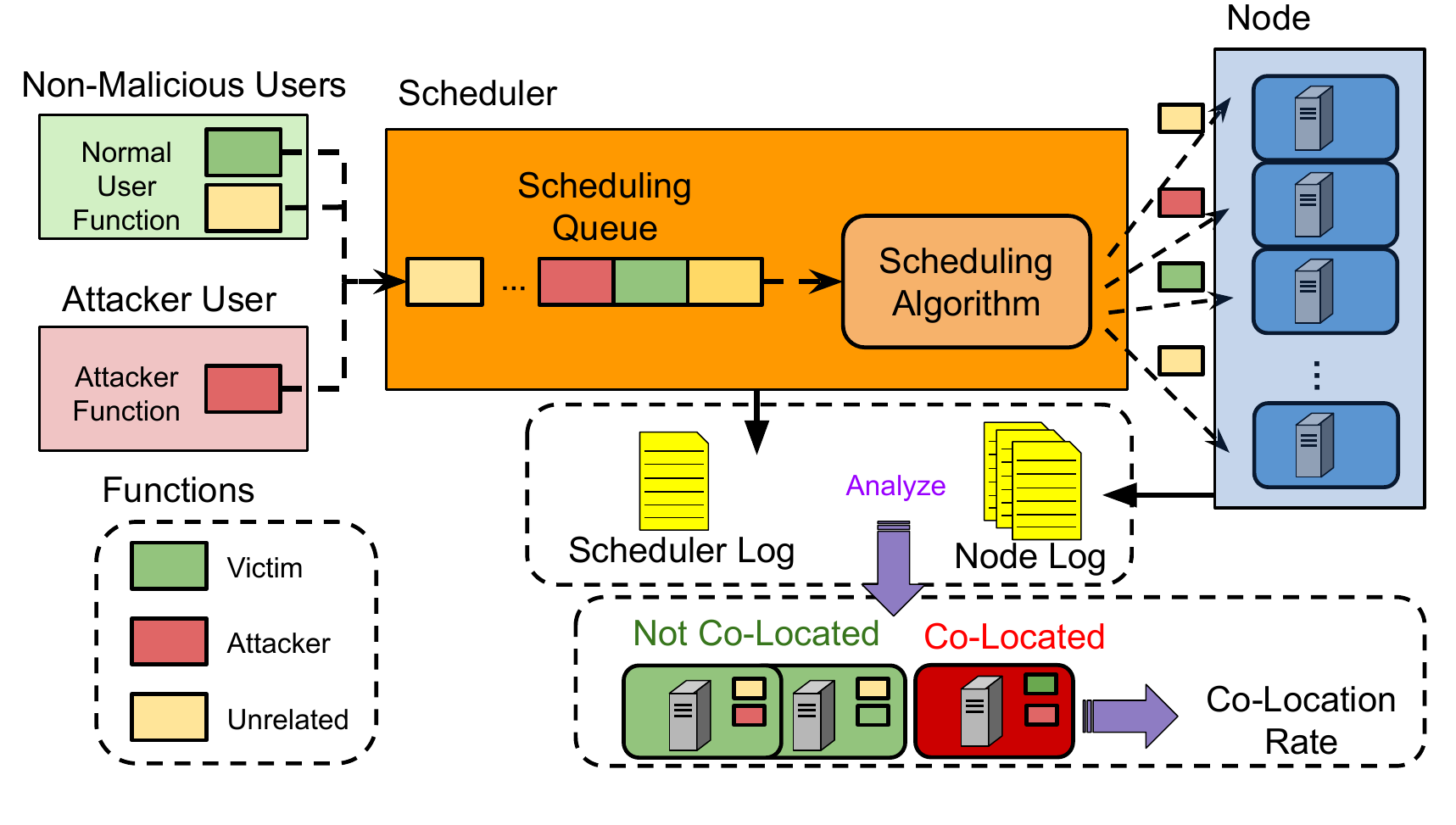}
    \caption{Architecture of our serverless system simulator.}
    \label{FigKumo}
\end{figure}

We conduct our fingerprinting experiments in a simulator written in C++. The usage of this simulator allows us to launch large-scale Monte Carlo experiments. The simulator consists of $\sim$2000 Line of Code and simulates the functionality of a serverless system, with a focus on the scheduler. There are $4$ different components in the simulator:
\begin{itemize}
    \item \texttt{Node}, which maintains node resource information;
    \item \texttt{Function}, which consists of attribute information of user-submitted functions;
    \item \texttt{Scheduler}, which models scheduling algorithms of different serverless platforms;
    \item \texttt{User}, which models user function invocation behaviors.
\end{itemize}
We implement these components as plug-and-play modules that interact with each other using uniform API interfaces, as shown in Fig.~\ref{FigKumo}.

In our design, the \texttt{User} modules are connected to the \texttt{Scheduler} module via a scheduling queue. \texttt{User}s will push \texttt{Function} objects to the scheduling queue, and \texttt{Scheduler} will consume the queue in a first-come-first-serve order. The scheduler will dispatch these function invocations to servers (\texttt{Node}) in the cluster using predefined scheduling algorithms. \texttt{Node} modules are responsible for maintaining server resource information and running function information. When a \texttt{Node} receives a \texttt{Function}, it will register the \texttt{Function} (if it is a cold-start invocation) and update the \texttt{Function}'s time-to-live (TTL) information. All \texttt{Function}s with TTL$=0$ will be eliminated from the system. \texttt{Scheduler} and \texttt{Node} modules will both produce log files for analysis.

At the beginning of the simulation, a list of users will be randomly generated, and an attacker will be initiated to target a specific user. Both non-malicious users and attackers will generate \texttt{Function} instances when they are initialized. The simulation will go through a pre-defined number of rounds of \texttt{Function} submissions by different users, and the log files will be collected for further analysis. We have implemented different scheduling algorithms in different \texttt{Scheduler} modules. 

\subsection{Selected Scheduling Algorithms}
We selected to implement the following schedulers in both the simulator and the Dask-based serverless platform:
\begin{enumerate}[label=\arabic*.]
    \item OpenWhisk~\cite{openwhisk}: OpenWhisk is an open-source serverless system. In this work, we select to use the \texttt{ShardingContainerPoolBalancer} algorithm provided in their repository~\cite{owrepo} as our target, as it involves an interesting invocation locality optimization.
    For each function to be scheduled, the scheduling algorithm selects a list of servers based on the hash value of the function. Every time the function is scheduled, the scheduler will examine the availability of servers on the list in order. This algorithm tries to balance cold-start reduction and load-balancing.
    
    \item Helper, which mimics the critical load-balancing behavior of Google Cloud Run~\cite{googlecloudrun} described in~\cite{zhao2024everywhere}. For each function to be scheduled, the scheduler selects a server as the base server first. As the load increases, the scheduler creates function host instances on other servers to perform auto-scaling.
    \item Package-Aware Scheduler (PASch)~\cite{aumala2019beyond}, which is a serverless algorithm that utilizes package locality to improve the performance of serverless platforms. The scheduler maintains a consistent hash ring~\cite{karger1997consistent} during operation and maps serverless functions to servers based on the largest package that is required to increase the package cache hit rate in the system and at the same time balance the load of each server in the system.
    \item Random, which is a scheduler that randomly dispatches functions to servers in the system.
\end{enumerate}

We utilize OpenWhisk~\cite{openwhisk} as an example of cold-start-optimized schedulers. Also, as it is one of the most popular open-source serverless platforms, our attack results will be representative. We selected Helper~\cite{zhao2024everywhere} since it is reported to be part of the scheduling algorithm of Google Cloud Run, and we also observe similar behaviors in Microsoft Azure (see Section~\ref{SecCase}). PASch is selected to represent schedulers with $\mathrm{F}_5$ (configuration-based locality) optimizations, and the attack methods on PASch can be easily expanded to target other schedulers with similar optimization features. Finally, Random is selected as the baseline and helps demonstrate if increased randomness can help defend against co-location attacks, as shown by~\cite{fang2023heteroscore} in serverful clouds. 

\subsubsection{Results}
We conduct this part of the experiments in simulation, in which we strictly follow the aforementioned submission strategies. The 3 phases of function invocations include:
\begin{enumerate}[label=Phase\arabic*., left=0cm]
    \item Invoking one function repeatedly;
    \item Invoking two identical functions repeatedly;
    \item Invoking multiple functions with identical functionality and different package requirements.
\end{enumerate}
Each scheduler processes a total of 8000 invocations per phase. The collected server ID traces are shown in Fig.~\ref{FigFinpTrace}. 
The discovered features of each scheduler are provided as follows:

\begin{figure*}[ht!]
    \centering
    \begin{subfigure}[t]{.36\linewidth}
         \centering
         \includegraphics[width=\linewidth]{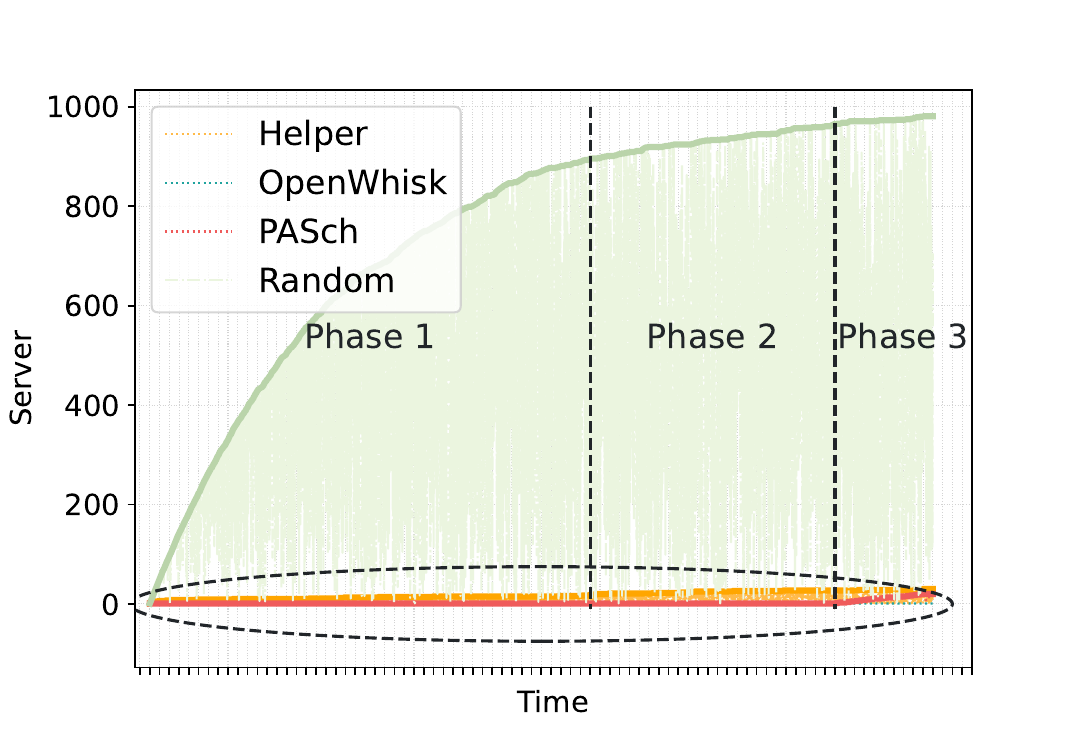}
         \caption{Full collected placement traces. }
    \end{subfigure}
    \centering
    \begin{subfigure}[t]{.62\linewidth}
         \centering
         \includegraphics[width=\linewidth]{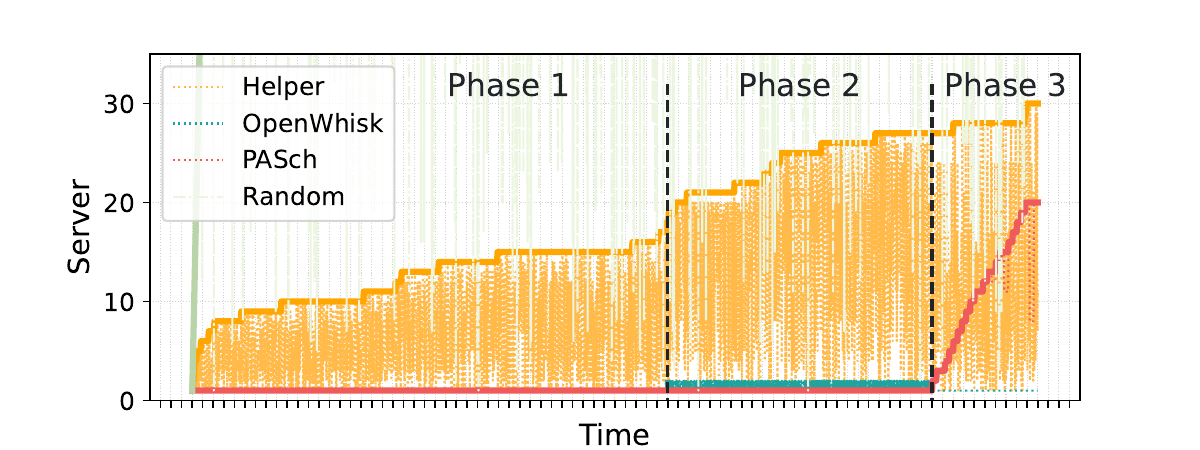}
         \caption{Zoom-in view of the circled area in (a).}
    \end{subfigure}
    \caption{Collected traces.}
    \label{FigFinpTrace}
\end{figure*}


\textbf{Random Scheduler.} 
It can be seen from Fig.~\ref{FigFinpTrace} 
that our function invocation sequence covers more servers than other types of schedulers, with 981 out of 1000 servers being covered in the cluster in our simulation. By more careful examination of the placement data, one can also find that the execution locations of serverless functions present no locality. These observations from the fingerprinting results match the features of the Random scheduler.

\textbf{Helper Scheduler.} 
By examining the placement trace in Phase 1, we can clearly observe that the scheduler has optimizations to reduce cold-starts, as functions are invoked repeatedly on the same set of servers ($\mathrm{F}_1$). As the number of function invocations increases, the scheduler starts to scale out function hosts by creating more instances to be placed on other servers ($\mathrm{F}_2$). In Phase 2, we can obtain the same observation as we start to involve a second function. By comparing the placement results of the overlapped function in Phase 1 and Phase 2, we find that if all hosts of a function are terminated, when re-launching these function hosts, the target scheduler will still choose the same set of servers, i.e., the Helper scheduler presents cold-start locality ($\mathrm{F}_3$). Also, we notice that the two different functions in Phase 2 are placed in different locations, indicating the Helper function does not confine the host placement to a small set of servers even though they are from the same user ($\mathrm{F}_4$). However, the behaviors of the Helper scheduler in Phase 3 are similar to Phase 1 \& 2, indicating that when making scheduling decisions, the scheduler does not consider package locality ($\mathrm{F}_5$).

\textbf{OpenWhisk Scheduler.} 
In Phase 1, when there is only one function involved, we can see that the function executions consistently stay on one machine, indicating that there is optimization to reduce cold-start ($\mathrm{F}_1$), yet no auto-scaling is involved ($\mathrm{F}_2$). In Phase 2, as we involve a second function, we can see that the number of covered servers increases by 1, indicating that function hosts from the same user are not bonded together ($\mathrm{F}_4$), and the location is the same when a function host experiences termination and cold-start ($\mathrm{F}_3$). In Phase 3, however, we do not notice the scheduling results change, indicating that the OpenWhisk scheduler does not use package information when making scheduling decisions.

\textbf{PASch Scheduler.} 
Similar to the OpenWhisk scheduler, in Phase 1, traces collected from PASch show instance locality ($\mathrm{F}_1$). However, since the placement consistently stays at the same machine, we can conclude that it does not perform auto-scaling based on the function invocation load ($\mathrm{F}_2$). In Phase 2, when a second function is involved, there is still only one server covered. This could potentially be due to (1) user locality, i.e., the scheduler places all functions from a user to a single machine; or (2) the identical configurations of the two functions result in the same placement decision. By examining Phase 3 of the trace, we can exclude possibility (1) since there are different function hosts in this phase. Besides, by comparing scheduling results in Phase 3, an attacker would be able to notice that the scheduler places function hosts with the same package requirements at the same node, demonstrating the existence of package locality-related policies in the scheduling algorithm.

\begin{table}[ht!]
\centering
\caption{A summary of the reverse-engineered features.}\label{TabScheFeatures}
\begin{tabular}{llllll}
\hline
Scheduler & $\mathrm{F}_1$         & $\mathrm{F}_2$         & $\mathrm{F}_3$         & $\mathrm{F}_4$         & $\mathrm{F}_5$         \\ \hline
Random    & \texttimes & \texttimes & \texttimes & \texttimes & \texttimes \\
Helper    & \checkmark & \checkmark & \checkmark & \texttimes & \texttimes \\
OpenWhisk & \checkmark & \texttimes & \checkmark & \texttimes & \texttimes \\
PASch     & \checkmark & \texttimes & \checkmark & \texttimes & \checkmark (package) \\  \hline
\end{tabular}
\end{table}

A summary of the exposed exploitable features is provided in Table~\ref{TabScheFeatures}. The above analysis results show that an attacker is able to decipher scheduler features based on the collected traces. The discovered features correctly match our example scheduler implementation, indicating that an attacker can obtain scheduler information simply by observing and comparing different parts of a collected trace. In reality, an attacker can flexibly perform logical reasoning based on the information contained in the collected traces and identify exploitable scheduler features.

\section{Recipe for Generating Co-Location Attacks}\label{SecAttack}
\subsection{Attack Construction Guide}
Based on the revealed scheduler features, an attacker can construct the attack method correspondingly and improve the efficiency of the co-location attack. Features identified by the previous scheduler fingerprinting step center around: (1) Policies that render the placement decisions more deterministic (locality); and (2) Policies that render function hosts to spread in the system (auto-scaling). These scheduling policies are related to the placement of function hosts and can thus be exploited by attackers to improve the attack success rate.

Our proposed attack consists of the following steps:

\textbf{Step 1.} We first examine $\mathrm{F}_4$ (Account Locality) from Section~\ref{SecScheFinp}. Schedulers employing this scheduling policy tend to place function hosts from the same account in a relatively confined set of servers, which limits the number of physical servers a single account can cover. This scheduling feature has been observed in AWS~\cite{wang2018peeking} and Google Cloud Run~\cite{zhao2024everywhere}. As a result, an attacker will need to employ multiple accounts to increase the chance of achieving co-location.

\textbf{Step 2.} In this step, we check if $\mathrm{F}_1$ (Invocation Locality) is employed and construct the attack accordingly. Warm-start-related optimizations are usually employed by cloud providers to avoid the costly cold-start process. Consequently, the placement of a single function will be limited to a small set of physical servers. Therefore, if feature $\mathrm{F}_1$ is identified, to improve attack efficiency, an attacker will need to construct multiple attack serverless functions.

\textbf{Step 3.} In this step, we explore how to exploit identified locality optimizations $\mathrm{F}_5$ (Configuration-Based Locality) from Section~\ref{SecScheFinp}. When this scheduling policy is revealed, an attacker can: 
\begin{enumerate}[label=\arabic*.]
    \item Focus on one specific choice of the corresponding scheduling parameter, and accurately target the victim's function. This occurs when the choice of scheduling parameter can be known by an attacker. For example, in a package-aware scheduler~\cite{aumala2019beyond}, when a victim hosts a popular serverless service like data analysis, the used packages are usually predictable. This tends to be the most efficient way of launching attacks, as shown by our results in Section~\ref{SecAttackEval}.
    
    \item Vary configuration parameters to enable the attack functions to spread in the cluster. For example, for package locality-aware schedulers~\cite{aumala2019beyond}, an attacker can construct serverless functions with different package dependencies to fully exploit the feature and force the corresponding function hosts to occupy different physical machines with a higher probability.
\end{enumerate}

\textbf{Step 4.} In Step 4, we observe whether $\mathrm{F}_2$ (Auto-Scaling) is involved. Auto-scaling tends to happen when a serverless function is in high demand. The scheduler will assign more function hosts in the system, causing the function hosts to spread across the cloud system. An attacker can exploit this feature to increase the physical server coverage of a single function by creating repeated invocation bursts, which is similar to the attack method used in~\cite{zhao2024everywhere}.

\subsection{Constructed Attacks for Selected Scheduling Algorithms}
Based on the discovered scheduler features shown in Table~\ref{TabScheFeatures}, we construct the following attack methods to tackle targeted schedulers:
\begin{enumerate}[label=M\arabic*., left=0cm]
    \item Targeting OpenWhisk, we construct an attack method that spawns multiple function invocations every time an attacker targets a victim. These serverless functions are of different names, and in reality, these functions can create hash collisions to enable co-location.
    \item Targeting the auto-scaling feature (Helper), in this attack strategy, attackers stress one of their own serverless functions by creating a burst of invocations, aiming to exploit the auto-scaling mechanism to achieve co-location.
    \item[M3-1] Targeting PASch, the attacker exploits the locality-based scheduling algorithm by creating functions with the same package dependencies as the victim function, increasing the likelihood of co-location with the victim function hosts.
    \item[M3-2] The attacker targets the same feature as M3 but varies the choices to increase coverage.
\end{enumerate}

We categorize M1, M2, and M3-2 as \textbf{scatter-based attacks}, since these attacks achieve co-location by scattering function hosts to increase the coverage, while M3-1 only targets the correct physical server. It is worth noting that M3-2 and M1 are different, although both involve creating multiple functions and deploying them. In M3-2, the attacker has identified the exploitable locality feature and can hence use this feature to scatter instances, while in M1, the attacker simply creates multiple copies of a single function, and all parameters remain the same.

\section{Attack Evaluation}\label{SecAttackEval}
\subsection{Serverless System Implementation}
Besides conducting experiments in the aforementioned simulator, we also deploy serverless cloud systems with different schedulers to a cluster. Due to resource and cost constraints, we choose to evaluate on a $50$-node cluster on CloudLab~\cite{duplyakin2019design} consisting of $50$ \texttt{r320} instances. Each node has one 8-core E5-2450 CPU and 16GB of Memory. We use one node as the central scheduler, and the others as workers.

The implementation of the serverless platform follows Abdi~\etal~\cite{abdi2023palette}. We implement the scheduling and function invocation functionalities entirely in the Dask Distributed framework~\cite{dask}, which is a Python-based distributed computing framework. Our implementation involves approximately 500 lines of changes across two files that together comprise around 16,000 lines of Python code. The changes are distributed as follows: modifications to \texttt{scheduler.py} for the scheduling algorithm implementation, and updates to \texttt{worker.py} to adapt the system for serverless operations. Functions are executed through the Azure Functions Host runtime~\cite{azurefunctionhost}. Upon cold-start, the server that is scheduled to execute the function will pull the function code from a remote destination and invoke the Azure Functions Host runtime. We use function benchmarks from benchmark sets such as FunctionBench~\cite{kim2019functionbench}. It is worth noting that this research focuses completely on the scheduler side; hence, we can safely omit environment configuration and execution details during the function execution process. We will use the log files produced by the scheduler node and worker nodes to obtain function placement information and calculate the required metrics. The four selected algorithms are implemented in this system.

\subsection{Metrics}
\textbf{Simulation Experiments.} The simulator enables us to flexibly change experiment parameters, e.g., cluster size, and conduct a large number of experiments. We launch 50 short-duration experiments in a 1000-node simulated cluster and use the percentage of experiments where the attacker successfully co-locates with the victim as our metric.

\textbf{Cloud Experiments.} Unlike in the traditional cloud instance placement situation, where instances are relatively fixed~\cite{fang2022repttack,fang2023heteroscore}, the flexible and scalable nature of serverless scheduling complicates the co-location measurements. Simply counting how many of the invocations to victim functions are co-located with attacker functions does not take the constant termination and start of instances into account. For example, an attacker's function invocation causes the corresponding function host to be placed together with a victim, resulting in subsequent invocations to this victim function being risky. The success of these subsequent invocations should be attributed to the specific function invocation that leads to co-location and shouldn't be counted multiple times.

We propose to use the following two metrics:
\begin{itemize}
    \item Attacker Efficiency (\textbf{AE}), i.e., the percentage of attacker function instance placement that results in co-location with victim instances;
    \item Placement Accuracy (\textbf{PA}), i.e., the percentage of victim instances that co-locate with attack instances.
\end{itemize}

\begin{figure}[ht!]
    \centering
    \includegraphics[width=\linewidth]{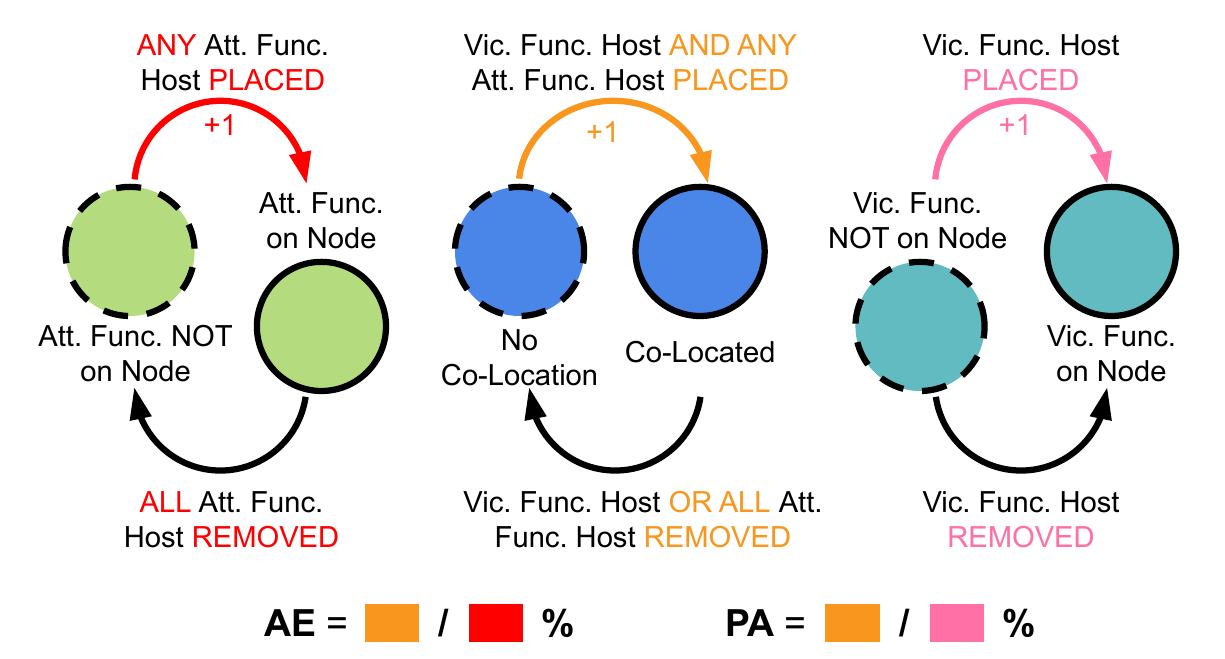}
    \caption{Diagrams of the AE and PA calculation process.  Each sub-diagram is a state machine, and we show the involved states and state transition conditions. To calculate AE and PA, we scan the collected log files and count the number of state transitions marked in \textcolor{transitionred}{red}, \textcolor{transitionorange}{orange}, and \textcolor{transitionpink}{pink}. These counts represent the number of attack instance placements, the number of times co-location happens, and the number of victim instance placements, respectively.}
    \label{FigMetric}
\end{figure}

The calculation process of AE and PA is shown in Fig.~\ref{FigMetric}. By counting state transitions marked as colored edges in Fig.~\ref{FigMetric}, we can obtain the number of function host placement activities that lead to co-location and the total number of attacker and victim placement activities.

\subsection{Results}
We perform the evaluation of our attack in both a simulation and a cluster. In the simulation experiments, the metric of focus is the short-term success rate, since it is easier to launch massive experiments in simulation. In each experiment, for a relatively short duration, we specify a victim user, and the attacker launches attack functions. We then collect the generated log files and calculate the percentage of successful experiments. In cluster experiments, we focus on a more realistic attack process with a longer duration and obtain the AE and PA data to measure the attack efficiency.

\begin{figure}[ht!]
    \centering
    \includegraphics[width=.8\linewidth]{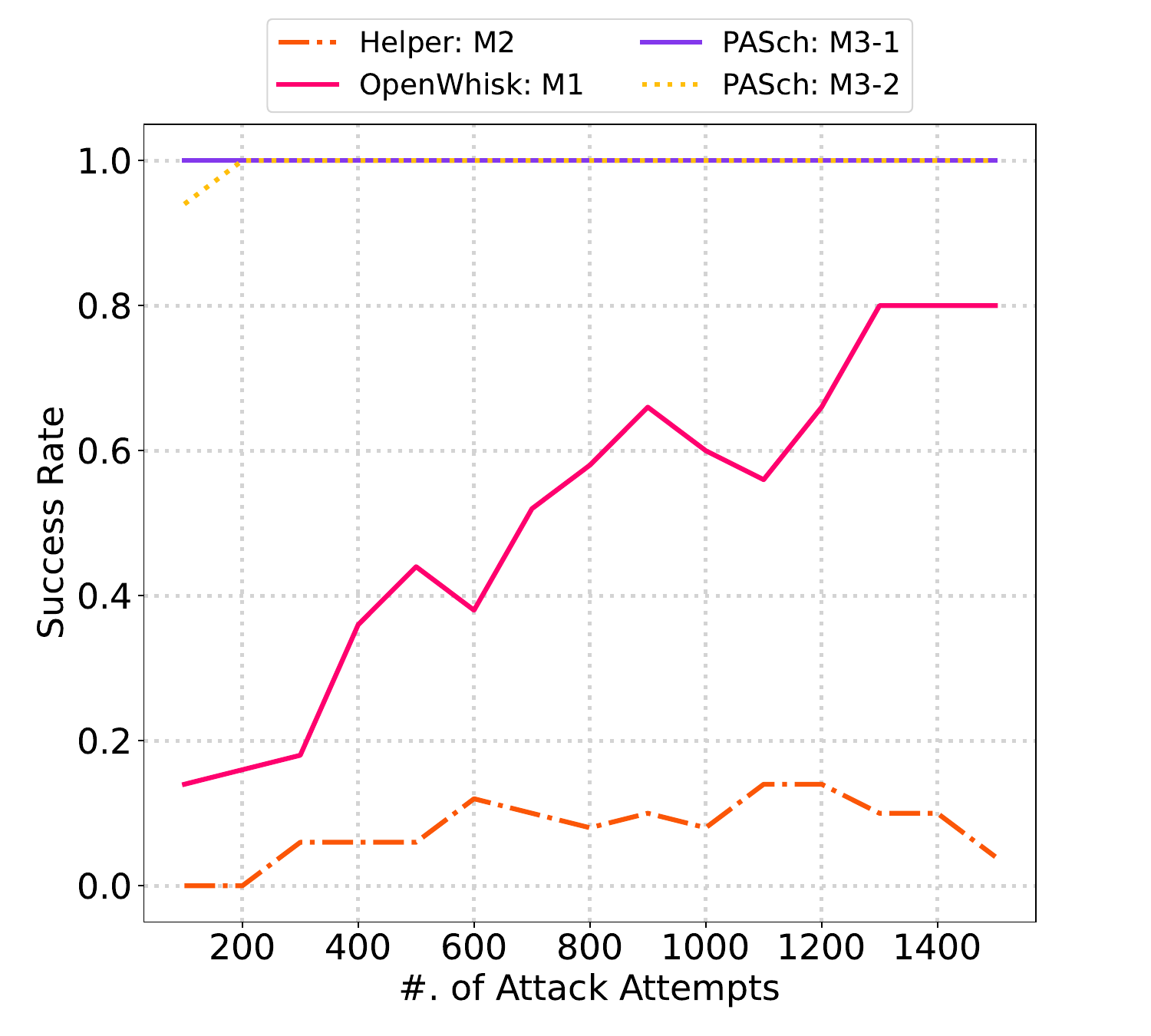}
    \caption{Short-term attack success rate obtained from the simulator.}
    \label{FigSimAttack}
\end{figure}

\textbf{Simulation.} The results we obtained from simulation are shown in Fig.~\ref{FigSimAttack}.

From Fig.~\ref{FigSimAttack}, we can see that in both PASch and OpenWhisk, the attackers are able to achieve relatively high co-location performance as they increase the number of attack functions. Also, it can be seen from the results of M3-2 that even if attackers do not possess full knowledge of the victim package dependencies, they can still achieve co-location by scattering their function instances. Below are some additional findings.

\begin{mdframed}[backgroundcolor=gray!20]
    \textbf{Finding 1}: Exploiting $\mathrm{F}_5$ (Configuration-based Locality) scheduling feature is efficient.
\end{mdframed}

By observing the two curves in Fig.~\ref{FigSimAttack} regarding PASch (the scheduler that considers package locality), we can see that the attacker is able to reach almost $100\%$ success rate. This indicates that as long as the locality feature of the scheduler is identified by the attacker, the attacker is able to easily achieve a high co-location success rate. If the user-submitted information is known by the attacker, e.g., the user utilizes certain packages, an attacker can precisely target the victim user's function host in the cluster by providing identical package requirements, which is the M3-1 attack strategy we provided in Section~\ref{SecAttack}. Otherwise, the attacker can construct multiple functions with various settings to ensure coverage, which is the M3-2 strategy. Either way, the attacker is capable of achieving a high success rate.

\begin{mdframed}[backgroundcolor=gray!20]
    \textbf{Finding 2}: Exploiting $\mathrm{F}_2$ (auto-scaling features) is not efficient.
\end{mdframed}

In Fig.~\ref{FigSimAttack}, as shown by the curve corresponding to Helper Scheduler (which has an auto-scaling feature), increasing the number of function invocations per attack attempt only slightly increases the success rate. This is due to the fact that not all function invocations can lead to the creation of a new function host. A more detailed theoretical analysis can be found in Appendix~\ref{AppTheo}. 

\begin{figure*}[ht!]
    \centering
    \begin{subfigure}[t]{.32\linewidth}
         \centering
         \includegraphics[width=\linewidth]{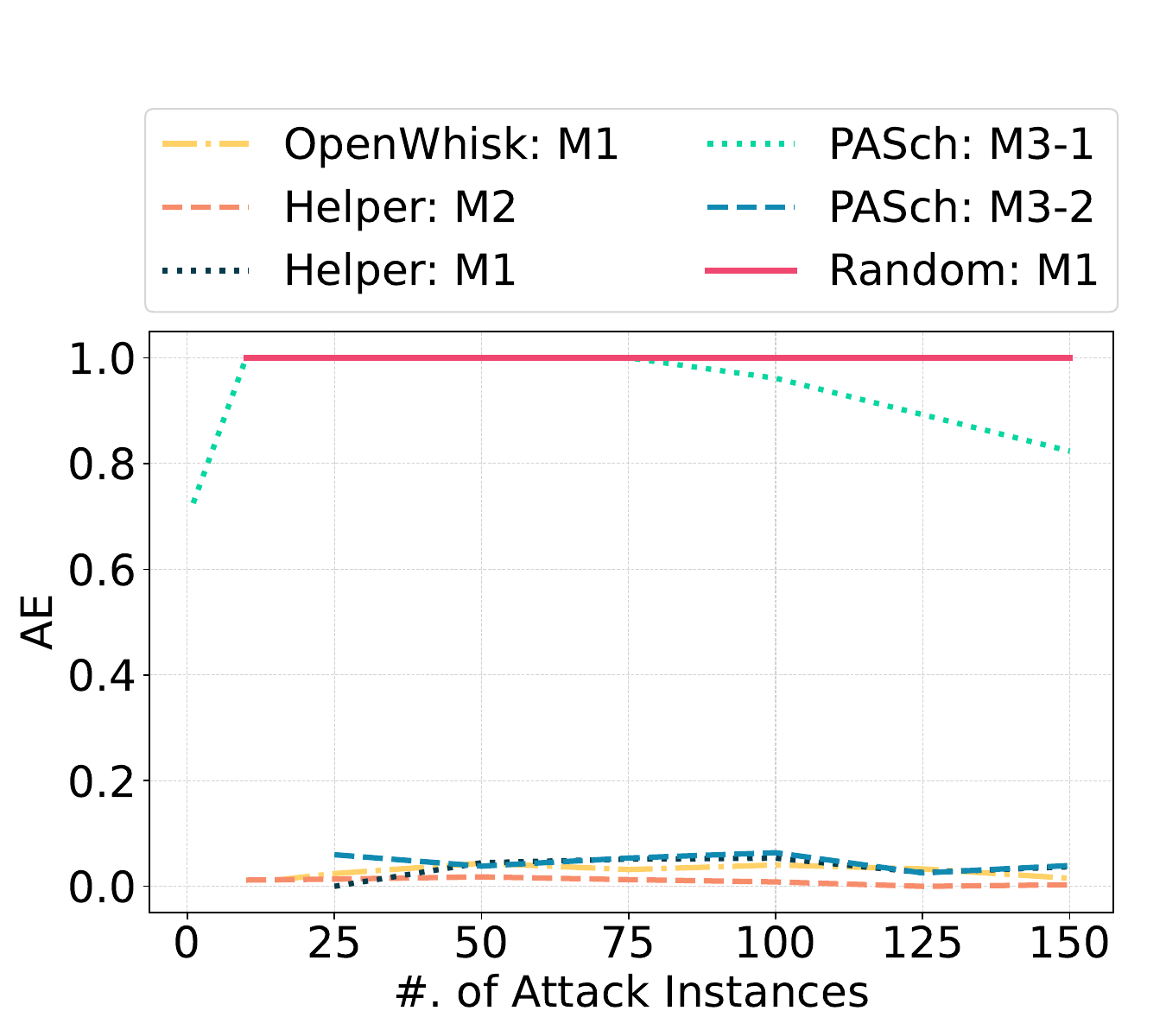}
         \caption{AE results.}
    \end{subfigure}
    \centering
    \begin{subfigure}[t]{.32\linewidth}
         \centering
         \includegraphics[width=\linewidth]{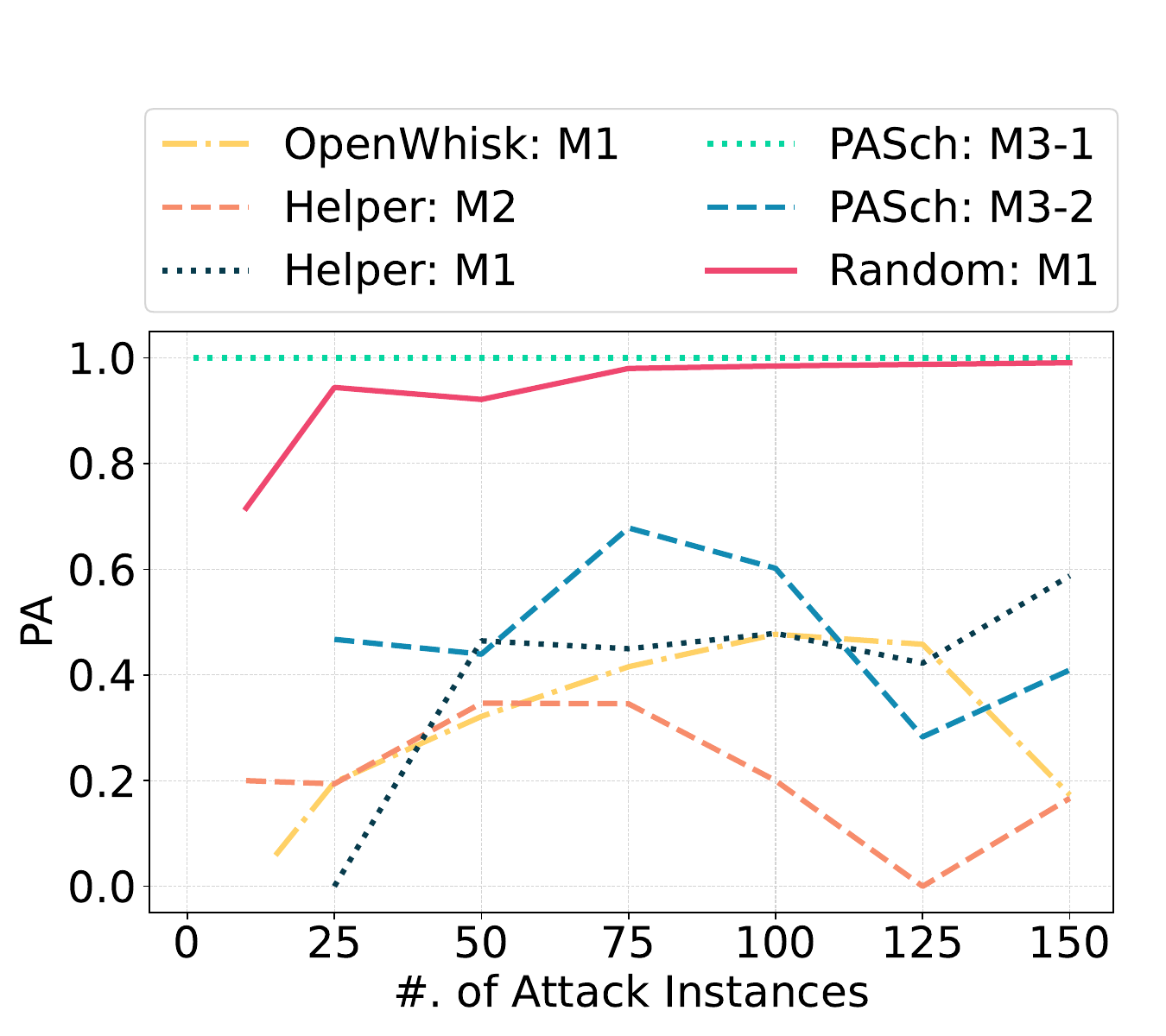}
         \caption{PA results.}
    \end{subfigure}
    \begin{subfigure}[t]{.32\linewidth}
         \centering
         \includegraphics[width=\linewidth]{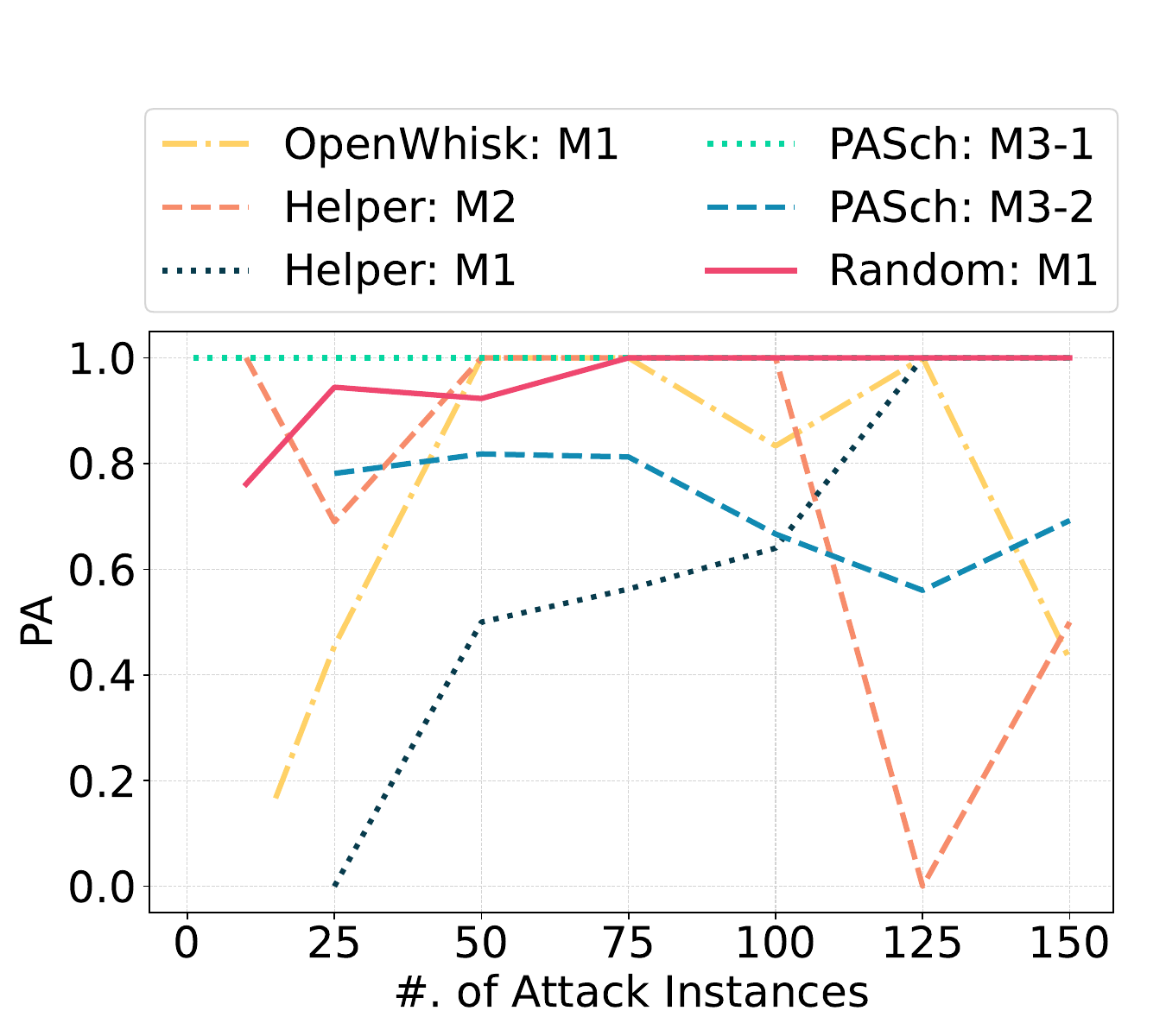}
         \caption{Maximum PA results.}
    \end{subfigure}
    \caption{AE and PA results collected from the cloudlab cluster.}
    \label{FigCluster}
\end{figure*}

\textbf{Cluster Experiments.} We also conduct experiments in a 50-node CloudLab~\cite{duplyakin2019design} cluster. 
All functions are invoked continuously during the experiments, with function invocation being submitted every two seconds. It takes 3-6 hours to run one experiment. We select to report AE and PA results of these experiments, which reflect the efficiency of the attacker's attempts and the effects of attacks on the victim function hosts. The results on the CloudLab cluster~\cite{duplyakin2019design} are shown in Fig.~\ref{FigCluster}. We can observe that the PASch scheduler and the Random scheduler have the highest PA values, indicating that most users' placed instances are co-located with a victim instance during its lifetime. By looking at the two curves corresponding to the Helper Scheduler (Helper: M1 vs Helper: M2), we can conclude that compared to using one attack function and repeatedly invoking to trigger the auto-scaling policies of the scheduler, using multiple functions results in higher PA values, i.e., more user function hosts are co-located with attack function hosts.

It is worth noting that, as we only specify a single victim and only repeat our experiments a few times, unlike the simulation experiments that are repeated hundreds of times, we can see random fluctuations in the results, which is close to the nature of this task in the real world. However, we are still able to see that with a relatively small number of attack functions, the attacker can achieve co-location with $\sim 50\%$ victim function hosts in most schedulers. 

We also observe that:

\begin{mdframed}[backgroundcolor=gray!20]
    \textbf{Finding 3}: Attack efficiency of scatter-based attacks (i.e. M1, M2, M3-2) is relatively low.
\end{mdframed}

From Fig.~\ref{FigCluster}~(a), we can see that most of the AE values of scatter-based attacks are below $0.1$ (OpenWhisk: M1, Helper: M1, and PASch: M3-2), meaning only a small portion of those placed attack function hosts achieves co-location with victim function hosts. The only exception is the Random scheduler, where the M1 attack strategy achieves an AE value of 1. 
An explanation of this is provided in Appendix~\ref{AppTheo}. Overall, our observations indicate that attacks would be more efficient if locality-optimization features of schedulers could be identified and the attacker could obtain related information about the victim.

The results in both the simulation and a cluster show that our constructed attack can enable attackers to effectively achieve co-location with victim function hosts. We also show that the most important thing an attacker can do is to identify scheduling locality features, as exploiting these features can greatly increase the success rate, as well as improve attack efficiency.

\begin{figure}[ht!]
    \centering
    \includegraphics[width=.75\linewidth]{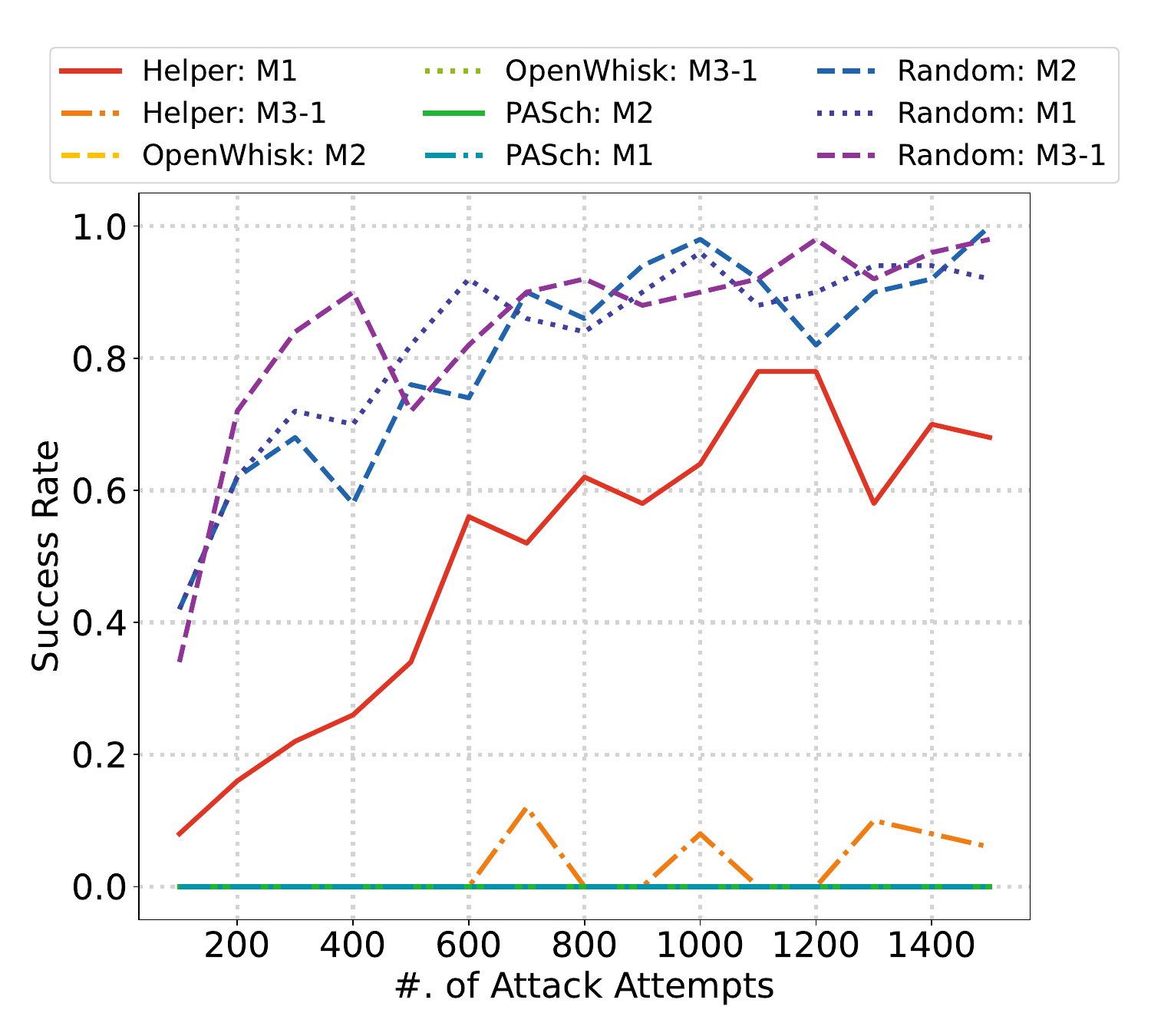}
    \caption{Attack transferability results.}
    \label{FigAttackTransfer}
\end{figure}

\textbf{Transferability.} We further examine whether the attack methods are transferable. We apply attack methods in Section~\ref{SecAttack} to mismatched types of schedulers. The experiments are conducted in the aforementioned simulator. The results are presented in Fig.~\ref{FigAttackTransfer}.

Fig.~\ref{FigAttackTransfer} shows that 4 out of the 9 curves (OpenWhisk: M2, OpenWhisk: M3-1, PASch: M1, and PASch: M2) consistently remain at $0$, indicating that the corresponding attack methods do not apply to targeted schedulers, even if he/she prepares a large number of functions to be deployed. This demonstrates the necessity of attackers to find the most appropriate attack method for a targeted scheduler. We have the following findings:

\begin{mdframed}[backgroundcolor=gray!20]
    \textbf{Finding 4}: Full randomness in scheduling is unsafe.
\end{mdframed}
We can see from Fig.~\ref{FigAttackTransfer} that all 3 attack methods (M1, M2, M3-1) on the Random Scheduler achieve similar performance, all better than other attack results obtained from mismatched schedulers and attack methods. This shows that randomness in the scheduling process can actually be used by an attacker to increase attack coverage.

\begin{mdframed}[backgroundcolor=gray!20]
    \textbf{Finding 5}: Using multiple functions to target schedulers equipped with $\mathrm{F}_2$ (auto-scaling) feature is effective.
\end{mdframed}
It can be seen from Fig.~\ref{FigAttackTransfer} that when facing schedulers with auto-scaling features (the Helper Scheduler), launching multiple different functions can enable attackers to reach a relatively high co-location success rate. Compared to results obtained from the Helper Scheduler in Fig.~\ref{FigSimAttack}, we can conclude that exploiting $\mathrm{F}_2$ (auto-scaling) feature with multiple functions is the more ideal solution.

\begin{mdframed}[backgroundcolor=gray!20]
    \textbf{Finding 6}: Locality-optimized schedulers are relatively safe when incorrect attack methods are used.
\end{mdframed}
The PASch~\cite{aumala2019beyond}-related curves in Fig.~\ref{FigAttackTransfer} show that it is hard to achieve co-location if attackers do not exploit the locality features. The success rates of both curves (PASch: M1 and PASch: M2) remain constant at 0. This is due to the fact that this type of scheduler tends to restrict the placement of functions to specific locations, and without targeting the correct scheduling features, the attacker will not be able to identify the victim's location.

Results from our simulation show that generally, without targeting correct scheduling features, an attacker will not be able to achieve co-location even with massive function invocations. The only exception is the scheduler with auto-scaling features, where it is more efficient to deploy multiple functions instead of simply triggering the auto-scaling policy to scale out. The results show that it is therefore important to identify exploitable features of schedulers, i.e., scheduler fingerprinting.

\section{Case Study: Microsoft Azure Functions}\label{SecCase}
As a case study, we also apply our fingerprinting method to Microsoft Azure Functions~\cite{microsoftazurefunctions}, which is one of the most prevalent serverless cloud platforms, and report our discovered scheduler features. Then we construct an attack accordingly and perform co-location attack experiments on Azure Functions. We create accounts in Azure Functions and utilize the Consumption Plan~\cite{consumptionplan} hosting model. In the Azure Functions Consumption Plan, users are billed based on cumulative memory resource usage, number of executions, and execution time. Users can only submit code to Azure, and are not allowed to deploy container-based applications in the Consumption Plan. This represents the least expensive and possibly the most common usage pattern of serverless platforms. All experiments are conducted using Consumption Plan~\cite{consumptionplan} in the \texttt{westus} region in Microsoft Azure.

\subsection{Scheduler Fingerprinting}
\begin{figure}[htb!]
    \centering
    \includegraphics[width=.8\linewidth]{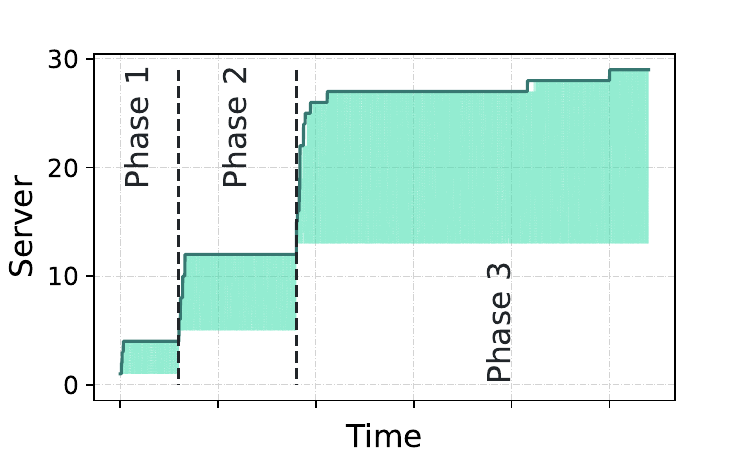}
    \caption{Fingerprinting traces collected from Azure Functions.}
    \label{FigFinpAzure}
\end{figure}

In Azure Functions, users can construct applications that contain multiple functions. In Phase 1 and Phase 2, we build function applications that only contain 1 function. In Phase 3, we also include function applications with multiple functions. We utilize the timestamp counter (TSC)-based method provided in~\cite{zhao2024everywhere} as our server fingerprinting method. This method reads the value of TSC registers inside CPUs, parses the base frequency of CPUs, and calculates the boot time as the watermark of a server. We construct an Azure Consumption Plan Function App by directly submitting the code directory, without using containers. We wait for at least 10 minutes after the experiment of each phase to allow function hosts to be terminated. The fingerprinting result sequences are shown in Fig.~\ref{FigFinpAzure}.

We have the following observations.

\begin{mdframed}[backgroundcolor=customblue]
    \textbf{Observation 1}: The scheduling algorithm of Azure Functions considers invocation locality ($\mathrm{F}_1$) and auto-scales if there is a high volume of incoming function invocations ($\mathrm{F}_2$).
\end{mdframed}

By examining the trace from Phase 1, we can see that the execution locations are relatively restricted, and these functions are executed at the same set of locations repeatedly. Also, in the experiment, as we provide a large number of function invocations per second ($\sim100$/s), we observe that the cloud scheduler gradually auto-scales function hosts to accommodate the large number of incoming function invocations per second. However, we are only able to increase the number of occupied physical servers to at most 4. 

\begin{mdframed}[backgroundcolor=customblue]
    \textbf{Observation 2}: In Azure Functions, function host cold-start locations are different ($\mathrm{F}_3$), and we do not observe account locality ($\mathrm{F}_4$).
\end{mdframed}

Compared to Phase 1, we can see that in Phase 2, the Azure Functions scheduler selects a completely different set of physical servers to execute function hosts, indicating that the function start locations are irrelevant to previous placement results. Besides, in Phase 2, the two constructed functions each cover 4 physical machines, showing that the scheduler of Azure Functions does not have the tendency to place different functions from the same user on the same physical machines.

\begin{mdframed}[backgroundcolor=customblue]
    \textbf{Observation 3}: Azure Functions places function hosts belonging to the same application on the same physical machine ($\mathrm{F}_5$).
\end{mdframed}

In Phase 3, we invoke multiple functions belonging to different Azure Functions applications. Some of these functions belong to the same function application. We observe that Azure Functions tend to place these functions on identical physical machines, even after auto-scaling. Meanwhile, functions belonging to different function applications are still placed on different machines. We were unable to observe other forms of $\mathrm{F}_5$ locality on Azure Functions.

\subsection{Co-Location Attack}
Based on the revealed scheduler features, we construct the following attack strategy:
\begin{mdframed}[backgroundcolor=customblue]
    \textbf{Attack}: 
    \begin{itemize}
        \item Use one attack account;
        \item Construct multiple attack functions belonging to different function applications;
        \item Create bursts of function invocations to trigger auto-scaling policies.
    \end{itemize}
\end{mdframed}

We create two accounts, one as the victim and the other as the attacker. The victim user runs one function in Azure Functions, which is our target. The attacker creates 64 different function applications, each containing one function. Both users utilize the Azure Consumption Plan, and the functions created by both users are the fingerprinting function, so that we can verify co-location. We deploy these functions, invoke them, and analyze the collected traces to determine if co-location is achieved.

\begin{figure}[ht!]
    \centering
    \includegraphics[width=.7\linewidth]{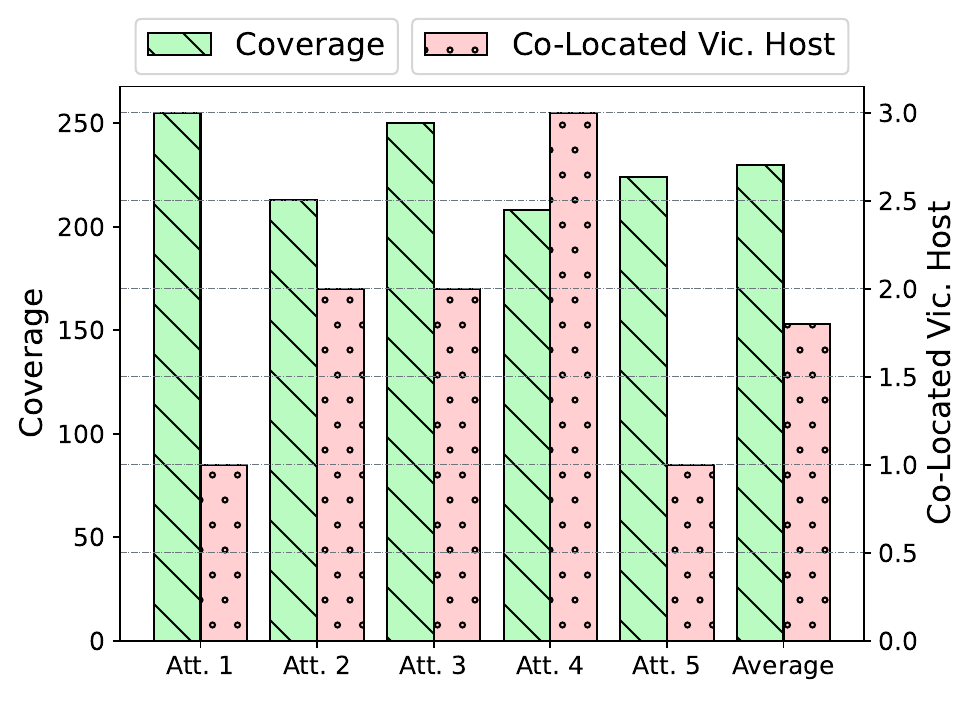}
    \caption{Co-location attack results on Azure.}
    \label{FigCoAzure}
\end{figure}

We launch the experiment 5 times. We observe that the attacker achieves co-location with at least one of the victim hosts in every experiment. Fig.~\ref{FigCoAzure} shows the results of the five co-location attempts. The 64 attack functions manage to cover $230$ machines on average, and on average $1.8$ victim function hosts are under attack. All the experiments, including fingerprinting and subsequent attack attempts, cost less than 25 USD combined from the attacker's side, indicating that it is not costly to achieve co-location with another user in the Azure Functions Consumption Plan. 

\section{Mitigation}\label{SecMit}
As demonstrated in our previous attacks, the vulnerability stems from the lack of enforced server isolation between users. Current scheduling algorithms enable attackers to easily reach servers hosting other users’ instances, either through instance spreading or precise targeting facilitated by scheduler locality optimizations. By intentionally minimizing overlap between servers occupied by different users in the scheduler’s design, the risk of co-location could be significantly reduced.
\subsection{\textit{Double-Dip} Scheduler}
We propose \textit{Double-Dip}, a scheduling algorithm that aims to provide ``soft'' server isolation for different users. The goal of \textit{Double-Dip} is to minimize the overlap of physical servers between different users. \textit{Double-Dip} works as follows:
\begin{enumerate}
    \item When a function $f_u$, belonging to user $u$, needs to be scheduled, the algorithm first checks if there exists a physical server $h$ such that:
    \begin{itemize}
        \item $u$ is already active on $h$, i.e. $u \in C(h)$, where $C(h)$ is the set of users served by $h$, and
        \item $h$ has sufficient resources to accommodate $f_u$ (i.e., $r(f) \leq R(h)$, where $r(f)$ is the resource requirement of $f_u$, and $R(h)$ is the available resources on $h$).
    \end{itemize}
    If such a server exists, assign $f_u$ to it. (This is where the name \textit{Double-Dip} comes from.)
    
    \item If no such host exists, the algorithm selects a physical server $h^*$ that:
    \begin{itemize}
        \item $h^*$ has the least variety of users (i.e., $h^* = \underset{h\in H}{\operatorname{argmin}} |C(h)|$), and
        \item $h^*$ has sufficient resources to accommodate $f_u$ ($r(f) \leq R(h^*)$).
    \end{itemize}
\end{enumerate}

The pseudocode of the \textit{Double-Dip} scheduling algorithm is provided as in Fig.~\ref{alg:spreadscheduler}.
\begin{algorithm}
\small
\caption{\textit{Double-Dip} Algorithm.}
\label{alg:spreadscheduler}
\begin{algorithmic}[1]
\Require $nodeList$: List of nodes (hosts), $appToRun$: Application (function) to be scheduled
\Ensure Application is scheduled on a node with minimal co-location rate increase.
\ForAll{$node \in nodeList$}
    \If{$\text{checkResource}(node, appToRun)$ \textbf{and} $\text{checkActiveApps}(node, appToRun)$}
        \State $node.\text{run}(appToRun)$ \Comment{Run the app on the same host with active apps}
        \State \Return $0$
    \EndIf
\EndFor

\Comment{No active apps found}
\While{true}
    \State $node \gets \text{FindLeastUserVarietyHost}(nodeList)$ \Comment{See Algorithm \ref{AlgFindLeastUserVarietyHost}}
    \If{$\text{checkResource}(node, appToRun.\text{resource})$}
        \State \textbf{break}
    \EndIf
\EndWhile

\State $node.\text{run}(appToRun)$ \Comment{Run app on the least user variety host}
\State \Return $0$
\end{algorithmic}
\end{algorithm}

Integrating \textit{Double-Dip} into commercial serverless clouds requires only minimal modifications to existing scheduling pipelines, as it operates entirely at the scheduling layer and leaves execution mechanisms unchanged. Major platforms such as AWS Lambda, Azure Functions, and Google Cloud Run already maintain per-function metadata; \textit{Double-Dip} simply leverages this to keep a lightweight active-user map per host and apply a user-aware rule during placement, and remains fully compatible with modern elastic environments. The scheduling overhead is small, consisting of constant-time user checks and a single pass to identify the least-user-variety host—operations already common in production load balancers. Our cost experiments further show that this logic scales well: even as we increase users and requests by an order of magnitude, warm-start behavior remains close to Helper and above OpenWhisk, with overhead saturating once workers have seen all function types. 

\subsection{Evaluation}
We conducted our experiments in a simulated environment to evaluate the performance of the \textit{Double-Dip} scheduling algorithm in a serverless cloud setting, comparing it with Helper. The simulation setup consisted of 100 nodes, each with the capacity to handle up to 1024 functions. The environment featured one victim function and a variable number of attackers, ranging from 5 to 50. Each attacker was programmed to perform 10 attack attempts during the simulation. To ensure statistical significance and reduce variability, each result was obtained by averaging data from 1000 independent simulation runs.

\begin{figure}[ht!]
    \centering
    \includegraphics[width=.8\linewidth]{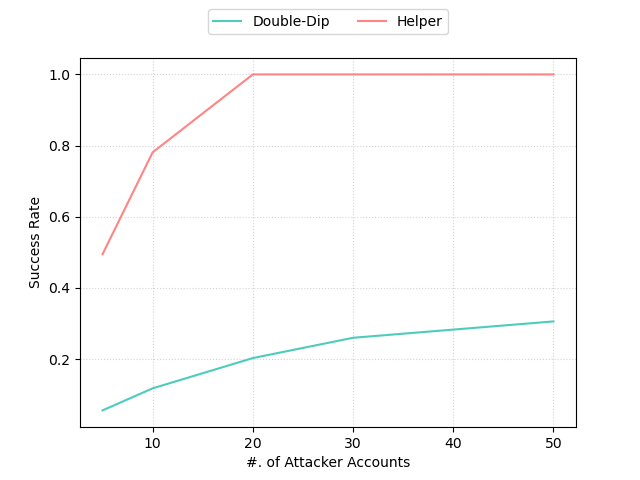}
    \caption{Co-location attack results on \textit{Double-Dip} and Helper}
    \label{FigDouble-Dip}
\end{figure}

The results are visualized in Fig.~\ref{FigDouble-Dip} with two lines, one representing our proposed scheduler and the other representing the Helper scheduler. The x-axis denotes the number of attackers, while the y-axis indicates the co-location success rate, where a lower success rate signifies better performance. At 5, 10, 20, 30, and 50 attacker accounts, our proposed scheduler achieved co-location success rates of 0.056, 0.118, 0.203, 0.260, and 0.306, respectively. In contrast, the Helper scheduler showed success rates of 0.495, 0.781, and 1.000 for attacker counts of 5, 10, and beyond.

These results demonstrate that \textit{Double-Dip} lowers the co-location rate an attacker can achieve even with multiple accounts, especially as the number of attackers increases. While the Helper scheduler consistently reaches a co-location success rate of 1.0 (indicating complete co-location success by attackers) for scenarios with 20 or more attacker accounts, the proposed scheduler maintains a much lower success rate, offering better resistance against co-location attacks.

\subsection{Cost Evaluation}
\label{subsec:warmstart}

To quantify the overhead of \textit{Double-Dip}, we measure the warm-start behavior of each scheduler under varying levels of multi-tenancy. We compare three schedulers: the OpenWhisk scheduler, the Helper scheduler, and our \textit{Double-Dip} scheduler. Similar to attack evaluation, we conduct experiments in a 50-node CloudLab~\cite{duplyakin2019design} cluster. Following the co-location experiments, we vary the number of users (5, 10, 20, 30, 50) and the number of total requests (500, 1000, 2000, 3000, 5000). All tasks are submitted concurrently, and warm-start ratios are computed from logs by tracking, for each worker, whether it has previously observed the corresponding function type.

\begin{figure}[ht!]
\centering
\includegraphics[width=.7\linewidth]{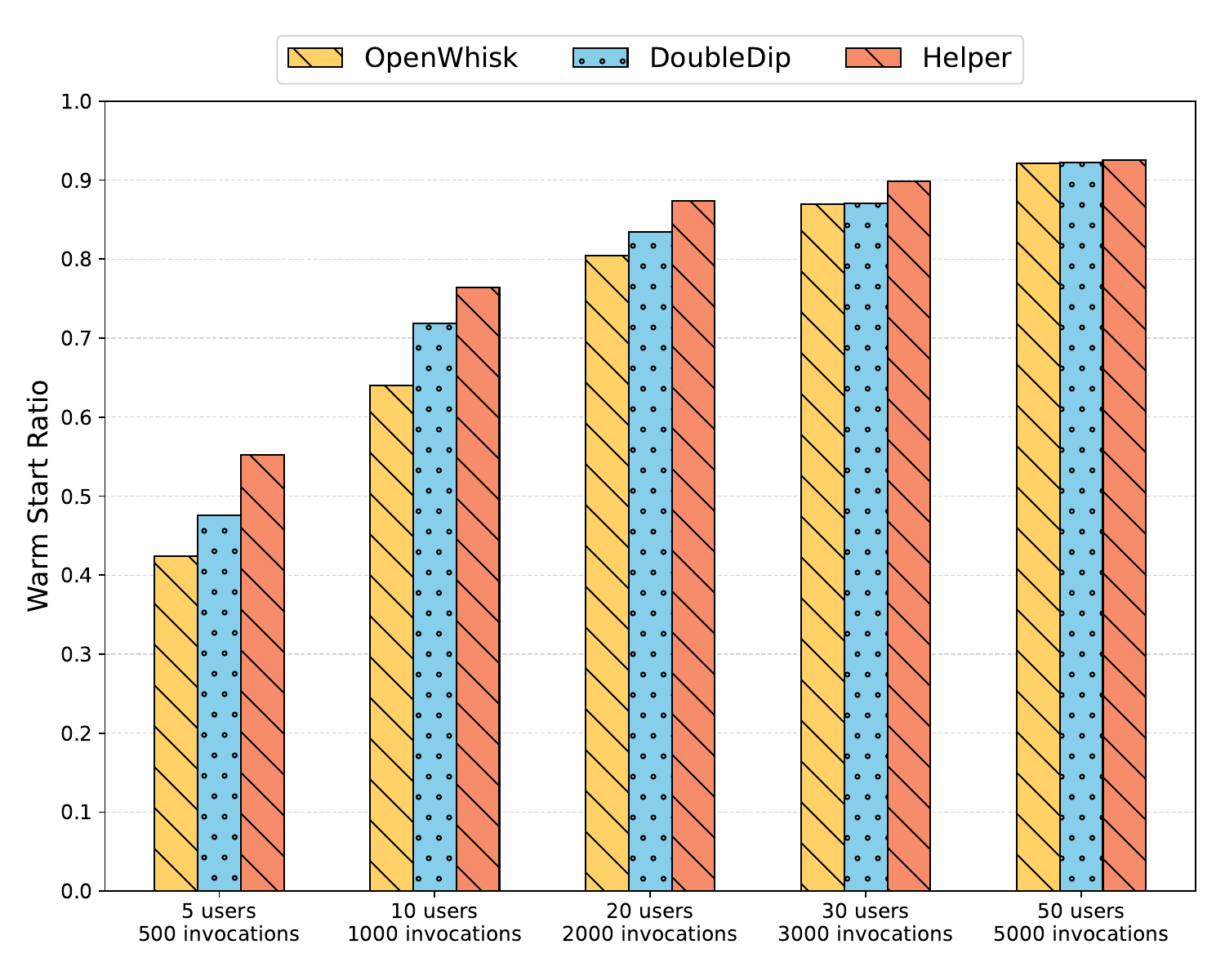}
\caption{Warm-start ratios.}
\label{fig:warm-start-ratios}
\end{figure}

Fig.~\ref{fig:warm-start-ratios} reports the warm-start ratios for all configurations. Across all settings, we observe that Helper consistently has the highest warm-start ratio, followed by \textit{Double-Dip}. This behavior is expected: Helper aggressively reuses recently active nodes, maximizing warm-starts but providing no user-level isolation. \textit{Double-Dip} lies in between, yet it still retains most of the warm-start advantages of Helper while offering substantially stronger isolation guarantees. The lower warm-start ratios in the smallest configuration reflect the system’s rapid scale-up, where new workers are activated quickly; this behavior is expected in serverless platforms and further highlights that \textit{Double-Dip} scales smoothly as the cluster expands.

We also observe that as the number of users and the total number of function invocations increase, the differences among the schedulers diminish; there is a 0.39\% difference between Helper and \textit{Double-Dip} at the 50 users, 5000 invocations configuration, indicating that \textit{Double-Dip} achieves lower relative overhead under high-intensity cloud-use scenarios.

Overall, \textit{Double-Dip} achieves warm-start performance that is close to Helper, while providing substantially stronger user isolation. These results indicate that \textit{Double-Dip} introduces only a modest performance cost.

\section{Discussion}\label{SecDisc}
\subsection{Security Implications in Scheduler Design}
Cloud scheduler is an important part of cloud systems. They are crucial to the performance of the whole system, and at the same time, the first barrier to cloud attackers. A good scheduler should optimize performance by fully exploiting function locality features while simultaneously ensuring unpredictability in scheduling results to enhance security. 

Our results help provide some insights into the security aspect of serverless scheduler design. On the one hand, we show that pure random schedulers are susceptible to co-location attacks, and attackers are guaranteed to achieve a high success rate as long as they prepare a sufficient number of attack functions to deploy. On the other hand, optimization policies that result in more deterministic instance placements, e.g., package-aware scheduling, can also be exploited by attackers to achieve co-location with victims. These results inspire a hybrid approach, where the scheduler still preserves the original optimization technology, which leads to determinism in instance placement, but at the same time adds randomness to the system to avoid the scheduling results being easily identified. The added randomness should not be excessive as well, otherwise the attacker would still be able to scatter attack function hosts and achieve co-location with victim hosts. It is important for the algorithm designer to strike the balance point and ensure a performant and secure system.

Besides, it is also important to provide user isolation. The co-location vulnerability in this paper largely originates from the fact that different users share the same scheduling pool, i.e., the whole cluster. If different users' scheduling pools are isolated or only partially overlapped, the difficulty of achieving co-location would significantly increase. This should not influence the performance of users' functions, since each function only scales out to a small number of physical servers even under a large burst of invocation requests, as shown by the results from Azure Functions. 

\subsection{Other Potential defense}
\label{Sec:OtherDefense}
Although the deployment of the \textit{Double-Dip} algorithm proposed in this paper can significantly reduce the co-location threat of serverless clouds, it does not provide a 100\% guarantee that the attacker and victim will not be co-located. Given the nature of serverless system design, i.e., it tries to abstract away cloud management details from users and increase server utilization rate from the vendor side, the only solution to guarantee instance isolation is to turn to traditional Infrastructure-as-a-Service (IaaS) clouds and reserve dedicated servers to use. However, this will significantly increase the cost. Below, we outline a ballpark estimation to compare the cost of using dedicated servers to provide an isolation guarantee and \textit{Double-Dip}.

Since the changes we propose in the \textit{Double-Dip} algorithm are oblivious to users, we assume the customers are still charged the same price as before. We calculate using the following settings:

\begin{enumerate}
    \item 30 Million requests per year (Coca-Cola’s vending machine service~\cite{cocacola});
    \item Each invocation requires one CPU core and 200ms to handle;
    \item Each function host is allocated 2GB of memory and 1GB of storage;
\end{enumerate}

In Table~\ref{TabPricing}, we provide a comparison of AWS and Azure. For AWS, we provide results from the US-East-1 region. We use the AWS Lambda Pricing Calculator~\cite{lambdapricingcalculator} to estimate the cost of serverless services, and their Elastic Compute Cloud (EC2) dedicated host pricing information~\cite{ec2dedicatedhostpricing} to estimate the cost of having a dedicated host. Similarly, for Azure, we use their dedicated host pricing list~\cite{azurededicatedhostpricing} and their pricing calculator~\cite{azurepricingcalculator} in the West US region.

\begin{table}[t]
    \centering
    \begin{tabular}{ccc}
    \hline
        Vendor & FaaS & Dedicated IaaS (Lower Bound)\\ \hline
        AWS~\cite{aws}    & \$$17.18$/mo  & \$$323.28$/mo \\
        Azure~\cite{azure}  & \$$33.3$/mo  & \$$444.03$/mo  \\
    \hline
    \end{tabular}
    \caption{Cost comparison of different service models.}
    \label{TabPricing}
\end{table}

We estimate the lower bound by using the lowest prices listed in the public pricing pages~\cite{ec2dedicatedhostpricing,azurededicatedhostpricing}. For AWS, we use the \texttt{a1} instance (0.449 USD/hour $\times$ 720 hour/month = 323.28 USD/month) to compute this lower bound. For Azure, we use the listed monthly cost of the \texttt{DCsv2-Type1} VM. We note that other host types can be significantly more expensive.

Using a dedicated host not only incurs substantially higher costs but can also lead to scalability limitations. For example, in AWS, a single dedicated host can accommodate only 16 \texttt{a1-medium} instances (based on the capacity information provided in~\cite{ec2dedicatedhostpricing}). This level of concurrency is insufficient to handle bursts of requests. In contrast, \textit{Double-Dip}, as a defense integrated within the serverless framework, preserves the platform’s ability to scale elastically and can therefore support applications with high and variable demand.

\section{Related Works}\label{SecRelatedWork}
\subsection{Serverless Cloud Scheduling}
Scheduler is considered one of the critical components of a serverless cloud system. In recent years, there have been works that target different features of serverless functions and propose corresponding optimizations to scheduling algorithms. Fuerst~\etal~\cite{fuerst2022locality} propose an algorithm named Consistent Hashing with Random Load Updates (CH-RLU), which improves the classic consistent hashing algorithm and takes server loads, function cold-start overheads, etc., into consideration to make trade-offs between server loads and function locality. Kaffes~\etal~\cite{kaffes2022hermod} offer a taxonomy of serverless scheduling algorithms and summarize a design parameter space of these schedulers. The authors then propose their algorithm, named Hermod, which significantly improves the performance of serverless systems. 

Regarding locality optimizations, Aumala~\etal~\cite{aumala2019beyond} propose PASch, which also uses consistent hashing and allows serverless systems to utilize the package locality of running functions to accelerate the function start process. Abdi~\etal~\cite{abdi2023palette} propose to use locality hints in schedulers that specify a function's preferences in data locality and optimize the performance of serverless functions. Li~\etal~\cite{li2022help} propose an algorithm to avoid function cold-starts in serverless systems by recycling idle function hosts of other functions and avoiding creating new function hosts.

\subsection{Co-Location Attacks}
Achieving co-location is an important and common prerequisite of various types of attacks. Most prior works target serverful clouds. The concept of co-location attacks in the cloud is first introduced and discussed by Ristenpart~\etal~\cite{ristenpart2009hey}. They launch malicious VMs in AWS EC2 and manage to carry out side-channel attacks. The authors achieve co-location by either issuing multiple attack VMs through brute force or exploiting the scheduling locality of timing. 
Han~\etal~\cite{han2015using} formalize the co-location problem and theoretically analyze the minimum number of VMs required under various scheduling strategies.  Makrani~\etal~\cite{makrani2021cloak} propose an attack method that targets machine-learning-based schedulers by disguising attackers' micro-architectural execution traces. Fang~\etal~\cite{fang2022repttack,fang2023heteroscore} discuss how to achieve co-location in a heterogeneous cloud and provide a method to quantitatively evaluate the co-location security threat.

The most relevant work is~\cite{zhao2024everywhere}. The authors propose a method to fingerprint physical servers and identify the auto-scaling feature of the Google Cloud Run platform, using which they manage to achieve instance co-location. However, compared to our work, their major focus is on the server fingerprinting step, and they only proved that their method works on one type of cloud scheduler. Our work focuses on the systemization of the co-location attack process, and our contribution is the construction of a universal attack strategy that can efficiently target different kinds of serverless cloud schedulers. Also, we propose a low-overhead mitigation scheduling algorithm, which is based on ``soft'' user isolation, to defend against co-location attacks.

\section{Conclusion}\label{SecConclusion}
In this paper, we focus on serverless cloud schedulers and propose methods to: (1) identify serverless scheduler policies; and (2) exploit revealed algorithm features to achieve serverless instance co-location. This series of techniques forms a practical guide for serverless cloud attackers, enabling them to identify scheduler vulnerabilities and construct targeted attacks. We perform thorough evaluations using simulations, a small-scale cluster, and Microsoft Azure Functions, covering schedulers in open-source platforms, commercial infrastructures, and experimental academic prototypes, proving the existence of co-location vulnerabilities in these serverless schedulers and the effectiveness of our attack. Future work will be dedicated to the design of more secure serverless scheduling systems.

\section{Ethics Considerations}
Our research has minimal impact on the public. The experiments conducted on Microsoft Azure involve only functions that read TSC register values, which neither interfere with co-located instances nor cause significant resource utilization. Additionally, the total duration of these experiments is limited to a few hours, ensuring negligible disruption to other users. All other experiments were conducted on our own devices or on servers provided by CloudLab~\cite{duplyakin2019design}, and therefore have no impact on the public.

The study of co-location itself does not directly compromise cloud users' instances. It would require subsequent attacks to cause real damage, which are beyond the scope of this work. We disclosed our findings to Microsoft, although they did not classify the issue as a vulnerability, noting that side-channel defenses are already in place. These defenses, however, are orthogonal to our work, which focuses on the scheduler side.

\bibliographystyle{IEEEtran}
\bibliography{mybib}

\appendices

\section{Fingerprinting Algorithm}\label{AppAlg}
The pseudocode is shown in Algorithm~\ref{AlgScheFinp}.
\begin{algorithm}
\caption{Scheduler Fingerprinting}\label{AlgScheFinp}
\begin{algorithmic}[1]
\Function{UpdateTrace}{}
\State \textbf{Input:} A server fingerprint $\zeta$ generated by $\mathcal{F}$
    \If{$\mathcal{F} \not\in \mathcal{R}$}  // A new server is discovered
        \State $\mathcal{R} \gets \mathcal{R} \bigcup \{\zeta \mapsto d\}$
        \State $d \gets d + 1$
    \EndIf
    \State $\mathbf{t} \gets (\mathbf{t},  \mathcal{R}(\zeta))$
\EndFunction

\Statex
\Statex
// Server fingerprint record, $\mathcal{R}(\zeta)$ returns a server ID
\State $\mathcal{R} \gets \{\}$
\Statex
// The trace sequence
\State $\mathbf{t} \gets ()_{0 \times 1}$
\Statex
// Index of the next discovered server
\State $d \gets 1$ 
\Statex

\Statex
// Phase 1.
\State Prepare a function $\phi_0$
\State $\Phi \gets \{\phi_0\}$    // Pool of constructed services
\For{a period of time}
    \State Invoke $\phi_0$
    \State \textsc{UpdateTrace}($\mathcal{F}(\phi_0)$)
\EndFor
\For{a period of time}
    \State Wait
\EndFor
\Statex

\Statex
// Phase 2.
\State Prepare $\phi_1$: a copy of $\phi_0$, and let $\Phi \gets \{\phi_0, \phi_1\}$
\For{a period of time}
    \State Select $\phi \in \Phi$ and invoke $\phi$
    \State \textsc{UpdateTrace}($\mathcal{F}(\phi)$)
\EndFor
\For{a period of time}
    \State Wait
\EndFor
\Statex

\Statex
// Phase 3.
\State Prepare $n$ variations of $\phi_0$ under the same name: $\phi_0'$, $\phi_0''$, ... 
\State $\Phi \gets \{\phi_0, \phi_0', \phi_0'', ... \phi_0^{(n)}\}$
\For{a period of time}
    \State Select $\phi \in \Phi$ and invoke $\phi$
    \State \textsc{UpdateTrace}($\mathcal{F}(\phi)$)
\EndFor
\State \textbf{return} $\mathbf{t}$

\end{algorithmic}
\end{algorithm}

\section{Theoretical Analysis of Attack}\label{AppTheo}
The co-location attack behaviors can be depicted using classical probability calculations. In this part, we provide theoretical analysis results of co-location attacks on different scheduler models with simplification. These analytical results enable an attacker to evaluate the cost of the attack.
\subsection{Basics}
In the rest of this section, we utilize the following notations:
\begin{itemize}
    \item Victim function: $v$;
    \item Attacker functions: $\{a_1, a_2, ...\}$;
    \item Cluster: $S = \{n_1, n_2, ..., n_{N}\}$.
\end{itemize}

We define the following events:
\begin{itemize}
    \item $A_i^\alpha$: After invoking the attack function $\alpha$ times, any of the attack function hosts are placed on node $n_i$;
    \item $V_i^\beta$: After invoking the victim function $\beta$ times, any of the victim function hosts are placed on node $n_i$;
    \item $C_i^{\alpha, \beta}$: Any of the aforementioned $\alpha + \beta$ invocations result in the co-location of victim and attacker function hosts on node $n_i$.
\end{itemize}

The indicator variable $I$ is defined as follows:
\begin{equation*}
    I(X) = 
    \begin{cases}
        1, &\text{if event $X$ happens} \\
        0, &\text{otherwise}.
    \end{cases}
\end{equation*}

The placement results of schedulers that consider only resource availability can be approximated as random events in a classical probability model. This can be explained as follows: since cluster information is agnostic to all users, from a user's perspective, the resource availability on physical servers is a random variable following the same distribution. This symmetry indicates that the probability of a user's instance being placed on any given node is equally $1/N$, where $N$ is the size of the cluster.

\subsection{Random Scheduling}
We first consider the scenario where every invocation to functions is randomly dispatched to nodes in the serverless cluster. In this scenario, each function invocation is considered independently, and no locality optimization is taken. If the assigned machine does not have a corresponding warm function host running, it results in a cold-start process. This scheduling policy is load-balanced but suffers from low performance due to the lack of locality optimizations.

We first calculate the probability of event $A_i^\alpha$ and $V_i^\beta$. $\forall n_i \in S$: 
\begin{equation*}
    \begin{aligned}
       P(V_i^\beta) &= 1 - P(\text{None of the victim function hosts is on }n_i) \\ 
       &=1 - \dfrac{(N-1)^\beta}{N^\beta} \\
       &= 1-\left(1-\dfrac{1}{N}\right)^\beta.
    \end{aligned}
\end{equation*}
Similarly,
\begin{equation*}
    P(A_i^\alpha) = 1 - \left(1-\dfrac{1}{N}\right)^\alpha.
\end{equation*}
As the placements of victim and attacker instances are independent, we have:
\begin{equation*}
    \begin{aligned}
        P(C_i^{\alpha, \beta}) &= P(A_i^\alpha)P(V_i^\beta) \\
        &= \left[1 - \left(1-\dfrac{1}{N}\right)^\alpha\right]\left[1 - \left(1-\dfrac{1}{N}\right)^\beta\right].
    \end{aligned}
\end{equation*}
The expected number of servers that have co-located victim and attacker function hosts is
\begin{equation}\label{EqExp}
    \begin{aligned}
        E_C&=E\left[\sum_{i=1}^{N}I\left(C_i^{\alpha, \beta}\right)\right] \\
        &=\sum_{i=1}^{N}E\left[I\left(C_i^{\alpha, \beta}\right)\right]\\
        &= N\left[1 - \left(1-\dfrac{1}{N}\right)^\alpha\right]\left[1 - \left(1-\dfrac{1}{N}\right)^\beta\right].
    \end{aligned}
\end{equation}

From Eq.~(\ref{EqExp}) we can see that as the system keeps running, 
\begin{equation*}
    \lim_{\alpha,\beta \to \infty}E_C = N,
\end{equation*}
i.e., co-location tends to happen on almost every physical server if the number of invocations to victim and attacker functions is sufficiently large.

\subsection{Invocation Locality}
Now, let's consider the scenario where the scheduler introduces optimization to avoid cold-start and always assigns function executions to existing function hosts if there are any. In this scenario, we can define $\alpha$ as the number of different attack functions and $\beta$ as the number of launched victim function hosts. In this scenario, we have $\beta\equiv 1$. As there are $\alpha$ different functions, there will be $\alpha$ different function hosts that are placed on physical machines in the cluster.

We can calculate the probability of co-location in the system as follows:
\begin{equation*}
    \begin{aligned}
        P_C=P(\text{Co-Location}) &= \dfrac{\sum_{i=1}^N N^\alpha - (N-1)^\alpha}{N^{\alpha + 1}} \\ 
        &= 1 - \left(1-\dfrac{1}{N}\right)^\alpha.
    \end{aligned}
\end{equation*}

By substituting $\beta=1$ in Eq.~(\ref{EqExp}), we have
\begin{equation*}
    E_C = 1 - \left(1-\dfrac{1}{N}\right)^\alpha.
\end{equation*}

As the attacker invests more and increases $\alpha$, 
\begin{equation*}
    \lim_{\alpha\to \infty}P_C=\lim_{\alpha\to \infty}E_C=1.
\end{equation*}

\subsection{Auto-Scaling}
Now let's consider the scenario where the scheduler additionally applies auto-scaling policies when there is a high demand for a function. The attacker stresses one attack function, enabling the attacker to cover more servers. 
Let $\alpha$ be the 
number of attack function invocations, and every $r$ invocations result in a new function host to be placed. We have
\begin{equation*}
        P_C=P(\text{Co-Location}) = 1 - \left(1-\dfrac{1}{N}\right)^{\frac{\alpha}{r}}.
\end{equation*}
We can see that since it requires $r$ invocations to spread a new function host, stressing the auto-scaling policy is not as efficient as having $\alpha$ different functions. In practice, an attacker should prioritize creating more functions, and stressing the auto-scaling policy can be an auxiliary strategy to increase coverage.
\subsection{Configuration-based Locality Optimizations}
Assume that additional locality features (e.g., package locality~\cite{aumala2019beyond}) are integrated into the scheduler. In this case, if the attacker provides the same specifications as the victim, he/she can increase the probability of having function hosts placed on the machine with a victim host, i.e.,
\begin{equation}
    P(A_i^1|V_i^1) = p > \dfrac{1}{N}.
\end{equation}
Assume the attacker prepares $\alpha$ different functions that exploit the locality feature. 
We have:
\begin{equation*}
    P_C = 1-(1-p)^\alpha >  1 - \left(1-\dfrac{1}{N}\right)^{\alpha}.
\end{equation*}
We can see that the exploitation of these features can increase the chance of co-location.

\section{Correctness of \textit{Double-Dip} Algorithm}
\subsection{Assumptions and Notation}
\begin{itemize}
    \item $H = \{h_1, h_2, \ldots, h_m\}$: Set of hosts.
    \item $f_u$: A function belonging to user $u$.
    \item $C(u)$: set of servers user $u$ occupies 
    \item $R(h)$: Available resources on host $h$.
    \item $r(f)$: Resource requirement of function $f$.
    \item Target: $min_{i,j}(|C(u_i)\cap C(u_j)|)$
\end{itemize}

\subsection{Claim: The algorithm minimizes co-location of users and respects resource constraints}
\textbf{Proof:}
We aim to show that the algorithm minimizes the co-location metric while respecting resource constraints. The proof proceeds by induction on the number of assignments made by the algorithm.

\textbf{Preliminaries:}
Let  $C(u)$ represent the set of servers user $u$ occupies, and let $C^*(u)$ denote the set of hosts after assigning function $f_u$ to host $h$. The co-location metric is defined as: 
\begin{equation*}
\sum_{i, j} \left| C^*(u_i) \cap C^*(u_j) \right|
\end{equation*}
and measures the degree of overlap in host assignments for all user functions. The algorithm assigns each function  to a host while ensuring that resource constraints are met.

\textbf{Inductive Step:}
\begin{enumerate}
    \item Case 1: If there exists a host \(h \in C(u)\) that has sufficient resources $r(f) \leq R(h)$, the function \(f_u\) is assigned to \(h\). In this case, the co-location metric remains unchanged, as the assignment does not introduce any new overlap:
    \[
    \forall i, j, \; \left| C(u_i) \cap C(u_j) \right| = \left| C^*(u_i) \cap C^*(u_j) \right|.
    \]
    Assigning \(f_u\) to any other host \(h \notin C(u)\) would unnecessarily increase the co-location metric. Thus, the choice of $h$ here is optimal, and the resource constraint is respected.

    \item Case 2: If no such host \(h \in C(u)\) exists, the algorithm selects a host $h^*$ that satisfies the resource constraint  and minimizes the increase in the co-location metric. Formally, the host  is chosen as:
    \begin{equation*}
        \resizebox{0.9\columnwidth}{!}{$
    h^* = \underset{h\in H}{\operatorname{argmin}} \left( \sum_{i, j} \left| C^*(u_i) \cap C^*(u_j) \right| - \sum_{i, j} \left| C(u_i) \cap C(u_j) \right| \right),
    $}
    \end{equation*}
    
    subject to:
    \[
    r(f) \leq R(h)
    \]

    Here, we select the host $h^*$ that minimizes the co-location rate increase by calculating the intersection of user co-locations. In the implementation, this is achieved by selecting the host with the least user variety, which corresponds to finding the host that will minimize the increase in the co-location rate. Algorithm~\ref{AlgFindLeastUserVarietyHost} is designed to identify such a host by evaluating the user diversity across nodes and selecting the one with the least variety. This approach ensures that the co-location rate increase, as described in the proof, is minimized.
\end{enumerate}
\begin{algorithm}
    \caption{findLeastUserVarietyHost Function}
    \label{AlgFindLeastUserVarietyHost}
    \begin{algorithmic}[1]
    \Require $nodeList$: List of nodes
    \Ensure A node with the least variety of users
    \State $minVariety \gets \infty$ 
    \State $selectedNode \gets \text{null}$
    \ForAll{$node \in nodeList$}
        \State $userSet \gets \text{getUsers}(node)$
        \State $variety \gets \text{size}(userSet)$ \Comment{Calculate the number of unique users}
        \If{$variety < minVariety$}
            \State $minVariety \gets variety$ \Comment{Update the minimum variety}
            \State $selectedNode \gets node$ \Comment{Update the selected node}
        \EndIf
    \EndFor
    \State \Return $selectedNode$
    \end{algorithmic}
    \end{algorithm}
\textbf{Greedy Property:}
At each step, the algorithm makes a locally optimal choice: either preserving the co-location metric ($h$) or minimizing its increase ($h^*$). This greedy approach ensures that the global co-location metric remains as small as possible after each assignment. The inductive hypothesis guarantees that all prior assignments are optimal with respect to the metric and resource constraints, and the current assignment extends this property.

\textbf{Conclusion:}
By induction, the algorithm minimizes the co-location metric globally while respecting resource constraints for every assignment. The proof accounts for all cases and guarantees that no alternative choice at any step would lead to a better solution.

\section{Validation of Server Fingerprinting Method}\label{AppTSC}
The TSC-based method was rigorously tested in~\cite{zhao2024everywhere}. However, their tests are conducted on Google Cloud Run~\cite{googlecloudrun} and need to go through a relatively complicated process. In this part, we validate that the TSC-based method also works on Microsoft Azure through relatively simple experiments.

Although this behavior is not officially documented, it is reasonable to assume that cloud providers co-locate functions accessing the same data to optimize locality and resource utilization. While co-location is not guaranteed, functions that share data, particularly those belonging to the same serverless application, are likely to have a higher probability of being physically co-located. We have the following assumptions:

\begin{enumerate}[label=\textbf{Assumption~\arabic*:}, leftmargin=*]
    \item Functions within the same application and access shared data are more likely to be co-located.
    \item TSC reads can be used as reliable server fingerprints.
\end{enumerate}

If we deploy function pairs $\{\texttt{func1}, \texttt{func2}\}_{(1,2,\ldots)}$ from the same serverless application that access shared data, and function pairs $\{\texttt{func1}’, \texttt{func2}’\}_{(1,2,\ldots)}$ from different serverless applications, then only when \textbf{BOTH} assumptions hold can we observe statistically significant differences in the TSC readings between functions from the same pair.

In our experiments, we deploy functions with TSC read capabilities. The function pairs are constructed under two conditions ($20$ pairs each): (1) both functions belong to the same serverless application and access shared data through a common Azure storage account; and (2) the functions belong to different serverless applications and do not access any shared resources. 

\begin{figure}[htbp]
    \centering
    \includegraphics[width=.8\linewidth]{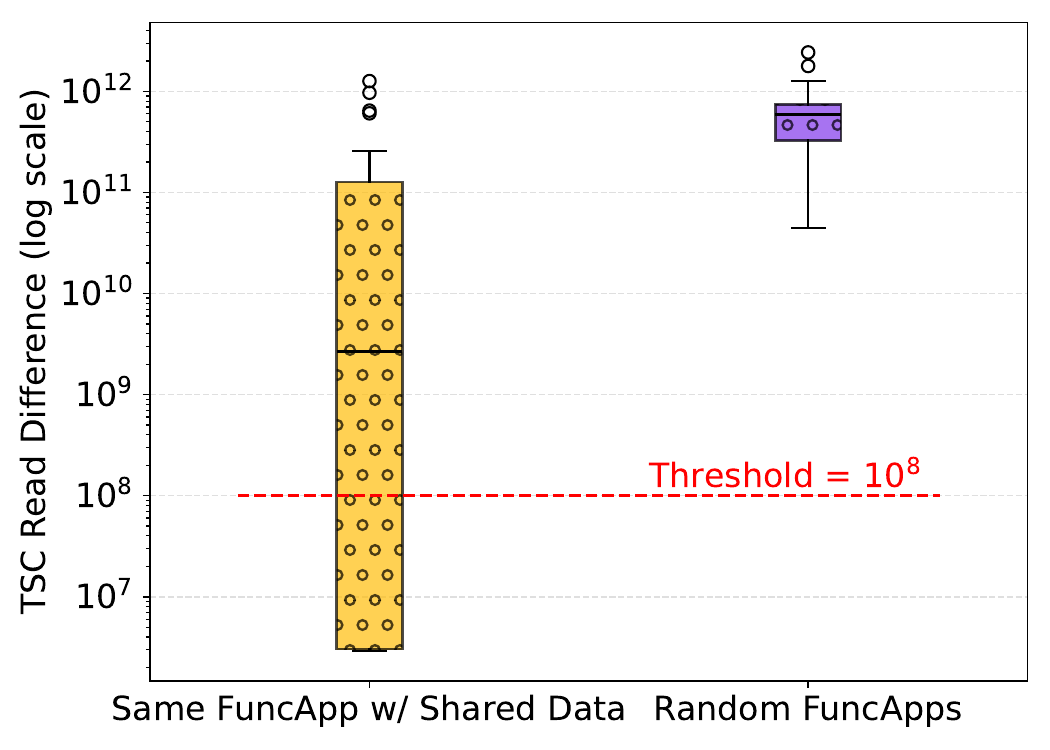}
    \caption{Comparison of TSC read differences between function pairs.}
    \label{FigTSCVal}
\end{figure}

The results are shown in Fig.~\ref{FigTSCVal}. As expected, the TSC read differences for condition~(1) (\textit{Same FuncApp w/ Shared Data}) are significantly lower than those for condition~(2) (\textit{Random FuncApps}); that is, function pairs likely to be co-located indeed exhibit similar TSC readings, whereas random function pairs show much higher TSC read variances. 

For accurate co-location measurement, we set the threshold to $t_h = 10^8$ (indicated by the dotted line in Fig.~\ref{FigTSCVal}). Any two measurements with a difference below this threshold are considered to originate from co-located function hosts.

\end{document}